# Three practical field normalised alternative indicator formulae for research evaluation[1]

*Mike Thelwall*, Statistical Cybermetrics Research Group, University of Wolverhampton, UK.

Although altmetrics and other web-based alternative indicators are now commonplace in publishers' websites, they can be difficult for research evaluators to use because of the time or expense of the data, the need to benchmark in order to assess their values, the high proportion of zeros in some alternative indicators, and the time taken to calculate multiple complex indicators. These problems are addressed here by (a) a field normalisation formula, the Mean Normalised Log-transformed Citation Score (MNLCS) that allows simple confidence limits to be calculated and is similar to a proposal of Lundberg, (b) field normalisation formulae for the proportion of cited articles in a set, the Equalised Mean-based Normalised Proportion Cited (EMNPC) and the Mean-based Normalised Proportion Cited (MNPC), to deal with mostly uncited data sets, (c) a sampling strategy to minimise data collection costs, and (d) free unified software to gather the raw data, implement the sampling strategy, and calculate the indicator formulae and confidence limits. The approach is demonstrated (but not fully tested) by comparing the Scopus citations, Mendeley readers and Wikipedia mentions of research funded by Wellcome, NIH, and MRC in three large fields for 2013-2016. Within the results, statistically significant differences in both citation counts and Mendeley reader counts were found even for sets of articles that were less than six months old. Mendeley reader counts were more precise than Scopus citations for the most recent articles and all three funders could be demonstrated to have an impact in Wikipedia that was significantly above the world average.

## 1  Introduction

Citation analysis is now a standard part of the research evaluation toolkit. Citation-based indicators are relatively straightforward to calculate and are inexpensive compared to peer review. Cost is a key issue for evaluations designed to inform policy decisions because these tend to cover large numbers of publications but may have a restricted budget. For example, reports on government research policy or national research performance can include citation indicators (e.g., Elsevier, 2013; Science-Metrix, 2015), as can programme evaluations by research funders (Dinsmore, Allen, & Dolby, 2014). Although funding programme evaluations can be conducted by aggregating end-of-project reviewer scores (Hamilton, 2011), this does not allow benchmarking against research funded by other sources in the way that citation counts do. The increasing need for such evaluations is driven by a recognition that public research funding must be accountable (Jaffe, 2002) and for charitable organisations to monitor their effectiveness (Hwang & Powell, 2009).

    The use of citation-based indicators has many limitations. Some well discussed issues, such as the existence of negative citations, systematic failures to cite important influences and field differences (MacRoberts & MacRoberts, 1996; Seglen, 1998; MacRoberts & MacRoberts, 2010), can be expected to average out when using appropriate indicators and comparing large enough collections of articles (van Raan, 1998). Other

---

problems are more difficult to deal with, such as language biases within the citation databases used for the raw data (Archambault, Vignola-Gagne, Côté, Larivière, & Gingras, 2006; Li, Qiao, Li, & Jin, 2014). More fundamentally, the ultimate purpose of research, at least from the perspective of many funders, is not to understand the world but to help shape it (Gibbons, Limoges, Nowotny, Schwartzman, Scott, & Trow, 1994). An important limitation of citations is therfore that they do not directly measure the commercial, cultural, social or health impacts of research. This has led to the creation and testing of many alternative types of indicators, such as patent citation counts (Jaffe, Trajtenberg, & Henderson, 1993; Narin, 1994), webometrics/web metrics (Thelwall, & Kousha, 2015a) and altmetrics/social media metrics (Priem, Taraborelli, Groth, & Neylon, 2010; Thelwall, & Kousha, 2015b). These indicators can exploit information created by non-scholars, such as industrial inventors' patents, and may therefore reflect non-academic types of impacts, such as commercial value.

A practical problem with many alternative indicators (i.e., those not based on citation counts) is that there is no simple cheap source for them. It can therefore be time-consuming or expensive for organisations to obtain, say, a complete list of the patent citation counts for all of their articles. This problem is exacerbated if an organisation needs to collect the same indicators for other articles so that they can benchmark their performance against the world average or against other similar organisations. Even if the cost is the same as for citation counts, alternative indicators need to be calculated in addition to, rather than instead of, citation counts (e.g., Dinsmore, Allen, & Dolby, 2014; Thelwall, Kousha, Dinsmore, & Dolby, 2016) and so their costs can outweigh their value. This can make it impractical to calculate a range of alternative indicators to reflect different types of impacts, despite this seeming to be a theoretically desirable strategy. The problem is also exacerbated by alternative indicator data usually being much sparser than citation counts (Kousha & Thelwall, 2008; Thelwall, Haustein, Larivière, & Sugimoto, 2013; Thelwall & Kousha, 2008). For example, in almost all Scopus categories, over 90% of articles have no patent citations (Kousha & Thelwall, in press-b). These low values involved make it more important to use statistical methods to detect whether differences between groups of articles are significant. Finally, the highly skewed nature of citation counts and most alternative indicator data causes problems with simple methods of averaging to create indicators, such as the arithmetic mean, and complicate the task of identifying the statistical significance of differences between groups of articles.

This article addresses the above problems and introduces a relatively simple and practical strategy to calculate a set of alternative indicators for a collection of articles in an informative way. The first component of the strategy is the introduction of a new field normalisation formula, the Mean Normalised Log-transformed Citation Score (MNLCS) for benchmarking against the world average. As argued below, this is simpler and more coherent than a previous similar field normalisation approach to deal with skewed indicator data. The second component is the introduction of a second new field normalisation formula, the Equalised Mean-based Normalised Proportion Cited (EMNPC), that targets sparse data, and an alternative, the Mean-based Normalised Proportion Cited (MNPC). The third component is a simple sampling strategy to reduce the amount of data needed for effective field normalisation. The final component is a single, integrated software environment for collecting and analysing the data so that evaluators can create their own alternative indicator reports for a range of indicators with relative ease. The methods are illustrated with a comparative evaluation of the impact of the research of three large medical funders using three types of data: Scopus citation counts; Mendeley reader counts; and Wikipedia citations.

## 2 Mean Normalised Log-transformed Citation Score

The citation count of an article must be compared to the citation counts of other articles in order to be assessed. The same is true for collections of articles and a simple solution would be to calculate the average number of citations per article for two or more collections so that the values can be compared. This is a flawed approach for the following reasons that have led to the creation of improved methods.

Older articles tend to be more cited than younger articles (Wallace, Larivière, & Gingras, 2009) and so it is not fair to compare averages between sets of articles of different ages. Similarly, different fields attract citations at different rates and so comparing averages between sets of articles from different mixes of fields would also be unfair (Schubert & Braun, 1986). One solution would be to segment each collection of articles into separate sets, one for each field and year, and only compare corresponding sets between collections. Although this may give useful fine grained information, it is often impractical because each set may contain too few articles to reveal informative or statistically significant differences.

The standard solution to field differences in citation counts is to use a field (and year) normalised indicator. The Mean Normalised Citation Score (MNCS), for example, adjusts each citation count by dividing it by the average for the world in its field and year. After this, the arithmetic mean of the normalised citation counts is the MNCS value (Waltman, van Eck, van Leeuwen, Visser, & van Raan, 2011ab). This can reasonably be compared between different collections of articles or against the world average, which is always exactly 1, as long as all articles are classified in a single field. If some articles are in multiple fields then weighting articles and citations with the reciprocal of the number of fields containing the article ensures that the world average is 1 (Waltman et al., 2011a).

A limitation of the MNCS is that the arithmetic mean is inappropriate for citation counts and most alternative indicators because they are highly skewed (de Solla Price, 1976; Thelwall & Wilson, 2016). In practice, this means that confidence limits must be calculated with bootstrapping and large sample sizes are needed for accurate results. An alternative approach that solves both of these problems is to switch from the arithmetic mean to the geometric mean because this is suitable for skewed data (Eysenbach, 2011; Fairclough & Thelwall, 2015b; Zitt, 2012). When the geometric mean replaces the arithmetic mean in the MNCS then the resulting geometric MNCS (gMNCS) does not have problems with skewed indicator data and confidence limits can be calculated with a mathematical formula (Thelwall & Sud, 2016). Alternative solutions for confidence intervals include stability intervals (Waltman, Calero-Medina, Kosten, Noyons, Tijssen, et al., 2012), although these use a type of bootstrapping.

An alternative solution for the field normalisation issue is to calculate the percentage of articles in the top X% (e.g., 1%, 10% or 50%) of the world separately for each field and year, then appropriately averaging the percentages across fields to give a single value (Schubert & Braun, 1996; Waltman & Schreiber, 2013). This is less precise overall than the geometric mean because it discards some of the citation information for any given percentile (Thelwall, 2016d). Choosing a set of percentiles to report (i.e., percentile rank classes: Bornmann, Leydesdorff, & Mutz, 2013) reduces this problem but generates multiple indicators.

A problem with the gMNCS is that the gMNCS world average is not guaranteed to be exactly 1 but can be slightly different. This is an unwanted complicating characteristic when combining the results of multiple fields. A simpler solution, proposed here (a variant of: Lundberg, 2007), is to log normalise citation or indicator data first, and then apply the standard MNCS formula. This allows the world average to be 1 without further calculations and introduces a single additional calculation step (see Appendix A for a worked example). Since citation counts and Mendeley readers for a single field and year typically conform reasonably closely to a discretised lognormal distribution (Fairclough & Thelwall, 2015b; Thelwall, 2016ab), a log transformation should result in data that approximately follow the normal distribution for these two. Although the discrete distribution can differ broadly in shape, it can be expected to be close enough to allow inferences based on the normal distribution assumption (e.g., Thelwall, 2016e). Because of this, it is reasonable to use the arithmetic mean on the log-transformed data and also to use standard formulae from the normal distribution to calculate confidence limits for both citations and Mendeley readers. For other alternative indicator sources if they can be proved to approximately follow the lognormal distribution then it would also be reasonable to use the normal distribution formula. In the case of data with high numbers of zeros, including most web indicators, the normal distribution formulae do not work well (see Appendix B for the Wikipedia citation data used in this paper) and so the assumptions below are not valid for them.

Since the logarithm is undefined at 0, the standard solution of adding 1 before applying the function is necessary. The formula for the resulting Mean Normalised Log-transformed Citation Score (MNLCS) for $n$ articles with citation counts or alternative indicator values $c_1, c_2, \ldots c_n$ is:

$$\text{MNLCS} = \left(\frac{\ln(1+c_1)}{l_1} + \frac{\ln(1+c_2)}{l_2} + \cdots \frac{\ln(1+c_n)}{l_n}\right)/n \tag{1a}$$

In the above formula, $l_i$ is the arithmetic mean of the $ln(1 + c)$ log-transformed world set of papers from the same field and year as $c_i$. Let $c_i' = \ln(1 + c_i)/l_i$ so that a simpler formulation is the following.

$$\text{MNLCS} = (c_1' + c_2' + \cdots c_n')/n \tag{1b}$$

This is similar to Lundberg's (2007) *item oriented field normalized logarithm-based citation z-score average*. Both formulae use the same $ln(1 + c)$ transformation and normalise the scores for individual articles with reference to the world average for a field and year. The difference is that Lundberg's formula normalises with $(\ln(1 + c_i) - l_i)/s_i$ instead of $\ln(1 + c_i)/l_i$ where $s_i$ is the standard deviation of the world log-transformed set for the field and year. Lundberg's formula therefore takes into account the variability in citation rates for individual fields and normalises for this to produce a difference indicator rather than a ratio indicator.

## 2.1 MNLCS confidence intervals, assuming accurate population data

Confidence limits can be calculated with the standard formula, where $t_{n-1,\alpha}$ is the two tailed critical value of Student's $t$ distribution with $n - 1$ degrees of freedom and confidence level $1 - \alpha$ and $s$ is the sample deviation of the log-transformed, normalised set $c_1', c_2', \ldots c_n'$.

$$\text{MNLCS}_L = \text{MNLCS} - t_{n-1,\alpha}s/\sqrt{n} \tag{2a}$$

$$\text{MNLCS}_U = \text{MNLCS} + t_{n-1,\alpha}s/\sqrt{n} \tag{2b}$$

For the large sample sizes typical of citation analysis, $t_{n-1,\alpha}$ is approximately 1.96 for a 95% confidence interval. This calculation assumes that the world averages used for each field/year are exact in the sense of being derived from the entire population, rather than from a sample of the population of interest, as discussed further in the paragraph below. If this is believed not to be true, then the methods of the next section can be used to calculate revised confidence limits.

In the social sciences it is standard practice to treat populations as being samples for statistical purposes, even though there is not universal agreement on this (Berk, Western, & Weiss, 1995; Bollen, 1995). The logic here is that social processes can never be precisely replicated and can be thought of instead as a sample of all possible outcomes from similar situations (for other justifications, see: Williams & Bornmann, 2016). The sample-based formulae are therefore recommended even when the entire population of articles is available.

## 2.2 Sampling methods

When calculating field normalised indicators, it seems to be standard practice to use the entire collection of articles from a given field and year, as recorded in the Web of Science (WoS) or Scopus, as the reference set for each article. This may be simple in practice for people with access to a complete set of citation counts from one of these databases but can be problematic for others that only have, or that can only afford, online access. Obtaining large sets of alternative indicator data can also be impractical when the values need to be calculated separately for each article rather than purchased in bulk from a data provider. For example, few webometric indicators (Thelwall & Kousha, 2015a) are currently available from data providers. The numbers involved can also be prohibitively large. If a medical funder wishes to evaluate a collection of its articles from the five previous years to identify trends, then these articles could span many different medical and health-related subject categories. The Scopus Medicine broad category includes 916,428 articles from 2015 alone [using the query SUBJAREA (medi) AND PUBYEAR = 2015]. For webometric indicators this number may be too large to be practical because of the time taken to gather the data and the need to pay for automatic Bing queries (see below).

The logical solution to this problem is to use sampling. For field normalisation purposes, this means calculating indicators for a random sample of the publications rather than for all of them. As long as the sample is large enough and random, or a close approximation to random, then replacing the full set with a random sample should not affect the results much, otherwise it will increase the size of the confidence intervals. Sampling would reduce precision more if the data were highly skewed, such as for raw citation counts, but this is not a problem here due to the log transformation. If the random sample for the world set has a standard deviation that is not much smaller than its mean, however, then the indicator confidence intervals can be affected more substantially. This is because the world set is used for the denominator of the indicator formula and if numbers close to zero are plausible for the denominator, then very high values are also plausible for the fraction as a whole. This can occur when the majority of the data values are zero, as for most webometric indicators and for citation counts from the current year, and so an alternative formula is needed for this case. This is dealt with in the section below.

## 2.3 MNLCS confidence intervals, not assuming accurate population data

Assuming that the world and group data are approximately normally distributed (see above), the normalised group values $c_i' = \ln(1 + c_i)/l_i$ in the MNLCS formula follow the ratio of two normal distributions, which is the Cauchy (or Lorenz) distribution. A confidence interval can be calculated for the Cauchy distribution using Fieller's theorem (Fieller, 1954; Motulsky, 1995, p. 285), although it is likely to be wide or undefined if the standard deviation of the world set is not small compared to the mean (see the discussion in the paragraph above). The formula for the lower and upper confidence limits $\text{MLNCS}_{FL}$ and $\text{MLNCS}_{FU}$ for the ratio of the group and world mean in the numerator and denominator of the $c_i^l$ values is as follows *for a single field and year*. Here, the confidence interval is undefined if $h > 1$. Also $SE_g$ and $SE_w$ are the standard errors and $\overline{c_g}$ and $\overline{c_w}$ are the arithmetic means of the normalised citations $\ln(1 + c_i)$ for the group and world sets, respectively.

$$h = \left( t_{n_1 + n_2 - 2, \propto} \frac{SE_w}{\overline{c_w}} \right)^2 \tag{3a}$$

$$SE_{\text{MNLCS}} = \frac{\text{MNLCS}}{1 - h} \sqrt{(1 - h) \frac{SE_g^2}{\overline{c_g}^2} + \frac{SE_w^2}{\overline{c_w}^2}} \tag{3b}$$

$$\text{MNLCS}_{FL} = \frac{\text{MNLCS}}{1 - h} - t_{n_1 + n_2 - 2, \propto} SE_{\text{MNLCS}} \tag{3c}$$

$$\text{MNLCS}_{FU} = \frac{\text{MNLCS}}{1 - h} + t_{n_1 + n_2 - 2, \propto} SE_{\text{MNLCS}} \tag{3d}$$

This method of calculating a confidence interval cannot be used if the data is taken from at least two different field/year sets because then the final ratio is a combination of at least two ratios, whereas the Cauchy distribution is for a single normal distribution in the ratio numerator and a single normal distribution in the denominator. In this case, the following heuristic approach is recommended.

1. Calculate the standard (2) and Fieller (3) versions of the confidence limits for each individual field/year set MNLCS.
2. Calculate the average width increase between the normal distribution (2) and Cauchy (3) formulae for each individual field/year. A weighted average should be used, with weights proportional to the number of articles in each field/year collection. Separate expansions should be calculated for the left hand side and right hand side of the confidence intervals since these are asymmetric.
3. Apply the standard formula (2) to the combined data and widen it by the average amount calculated as above.

In mathematical notation, the procedure is as follows. Suppose that the set of articles to be analysed $F = \{c'_1, c'_2, \dots c'_n\}$ is partitioned into $k$ different field/year subsets, $F_1, F_2, \dots F_k$. For each field/year subset $F_j$, let $\overline{c'_{F_j}}$ be the arithmetic mean of the normalised values in the $j$th subset $F_j$, and let $\text{MLNCS}_L(F_j)$ and $\text{MLNCS}_U(F_j)$ be the lower and upper 95% confidence limits calculated using the standard normal distribution formula (2). Similarly, let $MLNCS_{FL}(F_j)$ and $MLNCS_{FU}(F_j)$ be the lower and upper 95% confidence limits calculated using Fieller's method (3). Then the weighted average expansion rate from (2) to (3) can be calculated as follows.

$$AveExp_L(F_1, \dots F_k) = \frac{1}{n} \sum_{j=1,..k} |F_j| \left( \text{MNLCS}_L(F_j) - \text{MNLCS}_{FL}(F_j) \right) / (\overline{c'_{F_j}} - \text{MNLCS}_L(F_j)) \tag{4a}$$

$$AveExp_U(F_1, \dots F_k) = \frac{1}{n} \sum_{j=1,..k} |F_j| \left( MNLCS_{FU}(F_j) - \text{MNLCS}_U(F_j) \right) / (\text{MNLCS}_U(F_j) - \overline{c'_{F_j}}) \tag{4b}$$

Suppose now that the standard formula (2) is applied to the combined set $c'_1, c'_2, \dots c'_n$, with arithmetic mean $\overline{c'} = \text{MNLCS}$ giving a single set of 95% limits $\text{MNLCS}_L(F)$ and $MLNCS_U(F)$. Then the estimated MNLCS confidence limits can be calculated by widening these limits by the above weighted average.

$$\text{MNLCS}_{EL}(F) = \text{MNLCS}_L(F) - (AveExp_L(F_1, \dots F_k) + 1)(\bar{c}' - \text{MNLCS}_L(F))$$

(5a)

$$MNLCS_{EU}(F) = MNLCS_U(F) + (AveExp_U(F_1, \dots F_k) + 1)(MNLCS_U(F) - \bar{c}')$$

(5b)

In summary, Fieller's theorem can be used to generate confidence intervals for MNLCS values for individual field/year sets when samples are taken, and when multiple fields/years are involved then a heuristic formula (5) can be used to estimate the confidence limits. An alternative solution is to use a statistical bootstrapping approach to generate confidence intervals. When samples are taken, the world set should be bootstrapped as well as the group sets.

# 3   Equalised Mean-based Normalised Proportion Cited

If a data set for an indicator includes many zeros (i.e., uncited articles or articles with an alternative indicator score of 0) then, as argued above, the MNLCS confidence limits may be wide or undefined. A possible solution to this would be to remove all journals with too many zeros (Bornmann & Haunschild, 2016a) but this produces a biased subset of journals and may reduce the sample sizes substantially if there is a high overall proportion of zeros. A different approach is to calculate the proportion of articles that are cited or the proportion that have a non-zero indicator score. Focusing on the former case for terminological convenience, it could also be helpful to calculate a single combined proportion for each group.

Changing mathematical notation from the previous section, suppose that a group $g$ publishes $n_{gf}$ articles in field/year $f$ and $s_{gf}$ of them are cited. Suppose also that $F$ is the set of field/year combinations in which the group publishes. Similarly, let $n_{wf}$ be the number of articles that the world publishes in field/year $f$, with $s_{wf}$ of them being cited. The overall proportion of $g$'s articles that are cited is then the number of cited articles divided by the total number of articles.

$$p_g = \sum_{f \in F} s_{gf} / \sum_{f \in F} n_{gf}$$

(6)

If there are field/year differences in the proportion of cited articles and $g$ publishes different numbers of articles in each field/year then this proportion could be misleading. This is because if $g$ publishes more articles than average in fields with high proportions of cited articles then $p_g$ will tend to be larger.

For example, suppose that there are two medical humanities research groups, A and B, that produce research that is exactly average for the fields in which they publish, Medicine and Humanities. Within Medicine, 80% of the world's articles are cited and within Humanities, 20% of the world's articles are cited. Since Group A and Group B both produce sets of publications that are exactly average, 80% of both groups' Medicine articles are cited and 20% of both groups' Humanities articles are cited. The only difference between the two groups is that Group A publishes 200 articles in Medicine and 100 in Humanities, whereas Group B publishes 100 articles in Medicine and 200 articles in Humanities. Now the proportion of articles cited for group A is $(0.8 \times 200 \times 0.2 \times 100)/(100 + 200) = 180/300$ or 60%. In contrast, the proportion of articles for group B is $0.8 \times 100 \times 0.2 \times 200)/(200 + 100) = 120/300$ or 40%. Thus, Group A appears to be substantially better than Group B even though they both publish average research for their fields. The reason for this anomaly is that Group A publishes more research in the high citation field.

This problem can be avoided by artificially treating $g$ as having the same number of articles in each field/year combination, fixing this to be the arithmetic mean of numbers in each field/year. For this calculation, fields in which a group has published only a few articles should be excluded because these articles will have their importance inflated, although not in a biasing way if the group has the same citation differential in each field. As an initial heuristic, each field/year combination used should have at least 100 articles and should not be less than 25% of the mean. A more systematic and evidence-based approach is needed for this decision, however, based upon estimating the extent of additional variation introduced into the system for different group sizes and data distribution parameters.

Thus, for group $g$, the proportion of cited articles in field $f$ is still $s_{gf}/n_{gf}$ but when the different proportions are combined, $n_{gf}$ is replaced by $\hat{n}_g = \sum_{f \in F} n_{gf} / |F|$. Thus, the *equalised group sample proportion* $\hat{p}_g$ is the simple average of the proportions in each set, as follows:

$$\hat{p}_g = \frac{\sum_{f \in F} \frac{s_{gf}}{n_{gf}} \hat{n}_g}{\sum_{f \in F} \hat{n}_g} = \frac{\hat{n}_g \sum_{f \in F} \frac{s_{gf}}{n_{gf}}}{\sum_{f \in F} \hat{n}_g} = \frac{\sum_{f \in F} \frac{s_{gf}}{n_{gf}}}{|F|} \tag{7a}$$

The equalised world sample proportion has a similar formula.

$$\hat{p}_w = \frac{\sum_{f \in F} \frac{s_{wf}}{n_{wf}}}{|F|} \tag{7b}$$

The equalised group sample proportion has the disadvantage that it treats $g$ as if the average impact of its articles did not vary between fields. If $g$ had published more articles in fields/years in which it created higher impact articles (or at least more likely to be cited relative to the world average), then the equalised sample proportion $\hat{p}_g$ would be unfairly low. If, however, the assumption is made that $g$'s articles had the same level of impact relative to the world average in every field/year then the calculation would be an unbiased estimator. An alternative to this simplifying assumption is discussed in the section below.

Continuing the above example, Group A and Group B will both be treated as having 150 articles in each of Medicine and Humanities. In both cases the equalised group sample proportion is the same: $(0.8 \times 150 \times 0.2 \times 150)/(150 + 150) = 150/300$ or 50%. Thus Group A no longer gets an advantage for publishing more in a high citation specialism.

Because of the way in which the equalised group proportions are calculated, under the above assumption it is fair to compare $\hat{p}_g$ between groups and also against the world average $\hat{p}_w$. Confidence limits for these proportions can be calculated using Wilson's score interval (Wilson, 1927). These intervals would be optimistic if any of the group sample sizes $n_{gf}$ are much smaller than the others because their sampling deviations from the population proportion can be relatively large and exaggerated by the equalisation formula. For this reason, it is recommended to remove groups from the calculation when their sample sizes are small relative to the others.

The world Equalised Mean-based Normalised Proportion Cited (EMNPC) for each group $g$ is the ratio of the equalised sample proportions.

$$\text{EMNPC} = \hat{p}_g/\hat{p}_w \tag{8}$$

Now $\hat{p}_g/\hat{p}_w > 1$ implies that $g$ has a greater proportion of cited articles than the world average. This mirrors the situation for existing field normalised citation indicators, such as MNLCS, gMNCS and MNCS.

### 3.1 EMNPC confidence intervals

The ratio of two proportions is known as a risk ratio, and there are standard techniques for calculating lower $\text{EMNPC}_L$ and upper $\text{EMNPC}_U$ confidence limits in this case (Bailey, 1987). These assume that both the group and world sets are samples rather than populations. A continuity correction of 0.5 may be added to all $np$ terms.

$$\text{EMNPC}_L = \exp\left(\ln\left(\frac{\hat{p}_g}{\hat{p}_w}\right) - 1.96\sqrt{\frac{(n_g - \hat{p}_g n_g)/(\hat{p}_g n_g)}{n_g} + \frac{(n_w - \hat{p}_w n_w)/(\hat{p}_w n_w)}{n_w}}\right) \tag{9a}$$

$$\text{EMNPC}_U = \exp\left(\ln\left(\frac{\hat{p}_g}{\hat{p}_w}\right) + 1.96\sqrt{\frac{(n_g - \hat{p}_g n_g)/(\hat{p}_g n_g)}{n_g} + \frac{(n_w - \hat{p}_w n_w)/(\hat{p}_w n_w)}{n_w}}\right) \tag{9b}$$

Here, $n_g$ and $n_w$ are the combined group and world sample sizes, respectively, so that $\hat{p}_g n_g$ and $\hat{p}_w n_w$ in the formula are the number of group and world articles cited. A continuity correction (adding 0.5 to the number of cited articles for both the group and the world classes for the confidence interval width calculations) should be included in case the number of uncited articles is very small. These confidence intervals seem to be only approximations, however, as they can differ from bootstrapping estimates (Appendix C).

## 4   Mean-based Normalised Proportion Cited

As suggested by an anonymous referee, an alternative approach for calculating the proportion of articles cited is to echo the MNCS and, for each article, replace its citation count by the reciprocal of the world proportion cited for the field and year, if the citation count is positive, otherwise 0. Let $p_{gf} = s_{gf}/n_{gf}$ be the proportion of articles cited for group $g$ in field/year $f$ and let $p_{wf} = s_{wf}/n_{wf}$ be the proportion of the world's articles cited in field/year $f$. Then

$$r_i = \begin{cases} 0 & if \quad c_i = 0 \\ 1/p_{wf} & if \quad c_i > 0 \end{cases} \quad where\ article\ i\ is\ from\ field\ f \tag{10}$$

Following the MNCS approach, these zeros and reciprocals can now be averaged. This gives the same result as setting all citation counts that are greater than 1 to 1 and then applying the MNCS formula to the transformed binary data. This binary MNCS formula, or Mean-based Normalised Proportion Cited (MNPC), for the $n_g$ articles from group $g$ is therefore the arithmetic mean of the individual article values:

$$\text{MNPC} = (r_1 + r_2 + \cdots r_n)/n_g \tag{11}$$

This can be simplified to a weighted sum of the group ratio cited $p_{gf}$ to the world ratio cited $p_{wf}$ for all field/year combinations.

$$\text{MNPC} = \sum_{f \in F} \frac{n_{gf}}{n_g} \times \frac{p_{gf}}{p_{wf}} \tag{12}$$

There is a simple way to contrast MNPC and EMNPC, since (12) expresses MNPC as a sum of ratios, whereas from (8) EMNPC can be expressed as a ratio of sums:

$$\text{EMNPC} = \left(\sum_{f \in F} p_{gf}\right)/\left(\sum_{f \in F} p_{wf}\right) \tag{13}$$

This mirrors the difference between the new and old crown indicators (Waltman et al., 2011a) because the old crown indicator and EMNPC are ratios of sums whereas the new crown indicator and MNPC are sums of ratios. The old crown indicator was replaced by the new crown indicator partly because the old crown indicator effectively gave more influence to fields with higher average citation rates (Lundberg, 2007; Opthof & Leydesdorff, 2010; van Raan, van Leeuwen, Visser, & van Eck, & Waltman, 2010). This occurred because high citation fields could numerically dominate both the (single) numerator and (single) denominator of the old crown indicator. In the ratio of sums format above (13), it is clear that EMNPC inherits this disadvantage of the old crown indicator. For example, it gives an advantage to research groups that are particularly successful in fields with a high proportion of cited articles, because the other fields have less numerical influence in the EMNPC calculation. Suppose that groups A and B publish equal numbers of articles in fields X and Y with world average proportions cited 0.8 and 0.2, respectively. Suppose that A has 10% more articles cited than the world average in field X (i.e., 0.8x1.1=0.88) and 10% less in field Y (i.e., 0.18), whereas B has the opposite (0.72 cited in X but 0.22 in Y). Then the denominator for both is the same (1) and EMNPC for A is 0.88+0.18=1.06 whereas EMNPC for B is 0.94). In contrast MNPC for both is 1, which is fairer. Thus MNPC is intrinsically fairer than EMNPC, at least from a theoretical perspective.

The ratio of sums comparison (12, 13) also highlights the stability advantage of EMNPC for which it was created because a zero EMNPC denominator (13) is only possible if the denominators are zero for *all* fields. In contrast, the MNPC has a zero denominator (12) if there is a zero in *any* of the fields. This could be solved with the same approach as the new crown indicator by replacing the ratio 0/0 by 1 (Waltman et al., 2011a). No reasonable confidence interval could be calculated for this, however. When the denominator for a field includes the entire population then it is impossible to get a positive numerator and a 0 denominator since the numerator is based on a subset of the articles used for the denominator. When the MNPC denominator for a field is based on a sample rather than the whole set of articles published in the year and field then it is possible to get a positive numerator and 0 denominator (e.g., 1/0 and this occurs for the case study below) because the group set is not a subset of the world sample. It would not be reasonable to replace this infinite quantity with 1 and so MNPC could not be calculated without removing the field or replacing it with 1, both of which would be unfair because the group has a higher proportion cited than the world average. This makes MNPC intrinsically more fragile than EMNPC in the sense of either sometimes being undefined or unfair (when the sampling approach is used, which is likely to be standard practice in practical webometric applications) or confidence intervals to estimate its value being likely to be wider or undefined (when the sampling approach is *not* used, or when the sampling approach *is* used but no infinite ratio occurs). This advantage is more important for the EMNPC/MNPC context than for the new/old crown indicator context because MNPC and EMNPC were introduced to deal with the situation in which very low proportions of articles are cited and sampling is needed. Thus the fragility of MNPC can make it impractical for webometric indicators with a low proportion of cited articles.

## *4.1 MNPC confidence intervals*

There is not a simple formula for the MNPC confidence interval, but an approximate confidence interval can be constructed by taking the weighted sum of the confidence intervals for the individual risk ratios $p_{gf}/p_{wf}$ using the standard formula (Bailey, 1987). More specifically, if $\text{MNPC}_{fL}$ is the lower limit for group $g$ and if $\text{MNPC}_{fU}$ is the upper limit for group $g$ using the standard formula (Bailey, 1987) for field $f$ (see Appendix D), then the overall limits are:

$$\text{MNPC}_L = \text{MNPC} - \sum_{f \in F} \frac{n_{gf}}{n_g}\left(\frac{p_{gf}}{p_{wf}} - \text{MNPC}_{fL}\right) \quad (14a)$$

$$\text{MNPC}_U = \text{MNPC} + \sum_{f \in F} \frac{n_{gf}}{n_g}\left(\text{MNPC}_{fU} - \frac{p_{gf}}{p_{wf}}\right) \quad (14b)$$

A disadvantage of this formula is that if *any* of the constituent confidence intervals are large, then the overall confidence interval is also likely to be large, whereas this is not true for EMNPC since it comprises a single overall ratio. Moreover, if any of the world proportions are zero then the formula cannot be calculated unless the corresponding field is removed from all data, whereas EMNPC can always be calculated unless all world proportions are zero. In contrast to the EMNPC case, removing problematic sets is likely to bias the results since they would be removed for the potential to have very high or infinite values. These confidence intervals are only approximations, however, and can differ substantially from bootstrapping estimates (Appendix C).

# 5    Research Questions

This article primarily introduces a strategy for calculating field normalisation formulae and associated confidence intervals. The following research questions are designed to demonstrate the indicators rather than to give conclusive evidence of their value.

- RQ1: Are MNLCS, MNPC and EMNPC practical in sense of being able to distinguish between different groups?
- RQ2: Are EMNPC and MNPC preferable to MNLCS when the proportion of cited articles is low?

# 6    Data and Methods

Large medical research funders were selected to test the new formulae because these are important users of citation analysis and they fund research within a relatively narrow area. The National Institutes of Health (NIH) conducts and funds biomedical research in the U.S.A. The U.K. equivalent is the Medical Research Council (MRC). The U.K.-based Wellcome Trust biomedical research charity is the largest similar non-government research funder. All are based in advanced English-speaking nations and have similar remits and so are broadly comparable.

The three Scopus broad categories with the most articles overall for the three medical funders were chosen for the analysis: Medicine (MEDI); Biochemistry, Genetics and Molecular Biology (BIOC); and Immunology and Microbiology (IMMU). These three categories seem core to the work of the funders whereas other categories in which they have articles suggest a more peripheral contribution overall (e.g., Agricultural and Biological Sciences; Neuroscience). Any subject category is necessarily an oversimplification in the context of overlapping and evolving fields, as well as multidisciplinary articles. Alternative approaches have been proposed for categorising individual articles, including with the use of Mendeley data (Bornmann & Haunschild, 2016a; Haunschild & Bornmann, 2016).

Articles were analysed from 2013 to 2016 (the year of data collection) to allow an analysis of recent data and to assess the influence of time. The inclusion of recently published articles is important for many evaluations because recent research is likely to be the most relevant for practical applications and is important for web indicators (Priem, Taraborelli, Groth, & Neylon, 2010). Nevertheless, articles that were published early in a year have had relatively long to attract attention compared to articles published later in the year and this difference is most substantial for recent years. Hence, this can introduce a biasing factor if one of the groups analysed has published disproportionately few or many articles early or late in the year. Note also that for many scientometric evaluations, it is important to insist on only analysing articles that are old enough to have attracted sufficient citations to estimate their likely long term impact. Many bibliometricians recommend using citation windows of at least 3 years for useful results. Thus, the Scopus results for 2015 and 2016 are not relevant for many scientometric purposes.

For each year and category, all articles funded by each organisation were extracted from Scopus by searching for the funder name in the Scopus funder record. For each year and category, approximately 10,000 Scopus articles were also selected (the first and last 5000 published in each year) to form the world reference set (Tables 1-3). This is the maximum number of records that can be downloaded by querying Scopus for a complete list from a category. It is not a random sample but is balanced in terms of time and should not result in any systematic bias towards groups unless they tend to publish with a different temporal pattern than average for academia. Funding information is incomplete and sometimes incorrect in Scopus (Sirtes, 2013). The figures below may therefore cover only about a third of the journal articles funded by each source but since the information was collected in the same way for each organisation, this seems adequate for the purposes of comparing the methods. The three funders presumably have more complete lists of publications produced by their own databases and additional ad-hoc methods but, for testing the methods, gathering information in an identical way from Scopus seems to be reasonable. Assuming that the sample covers a third of the Scopus articles from each organisation then the confidence intervals would probably be $\sqrt{3}$ times narrower.

For the calculations, papers that occurred in multiple subject categories were counted as whole articles for all purposes. Thus, an article from any of the funders or the world set could have been included up to three times, once for each category. Although it might have given better results to weight each article by the reciprocal of the number of categories containing it (e.g., Waltman et al., 2012), this was not done here in order to keep the technique as simple as possible. Nevertheless, the fractional counting method should be used if it is clear that one group publishes its best (or worst) work in an unusually high proportion of multiply-classified (or single classified) fields.

**Table 1**. Descriptive statistics for the Scopus citation data. Each set contains three fields: Medicine; Biochemistry, Genetics and Molecular Biology; and Immunology and Microbiology.

| Sample Mean Nonzero | 2013 | 2014 | 2015 | 2016 |
|---|---|---|---|---|
| World | 29928 6.0 71% | 29952 3.3 72% | 29875 1.1 42% | 29950 0.20 10% |
| MRC | 1695 14.2 94% | 3410 8.9 90% | 2231 3.4 75% | 783 0.36 21% |
| NIH | 22571 13.1 95% | 24447 7.7 91% | 23489 2.7 67% | 14212 0.30 19% |
| Wellcome | 1950 14.4 96% | 2208 11.1 91% | 1363 3.7 72% | 535 0.30 19% |

**Table 2**. Descriptive statistics for the Mendeley readership data. Each set contains three fields: Medicine; Biochemistry, Genetics and Molecular Biology; and Immunology and Microbiology.

| Sample Mean Nonzero | 2013 | 2014 | 2015 | 2016 |
|---|---|---|---|---|
| **World** | 29928 12.4 69% | 29952 10.5 77% | 29875 7.0 70% | 29950 2.8 55% |
| **MRC** | 1695 30.8 96% | 3410 27.3 95% | 2231 19.4 91% | 783 8.7 81% |
| **NIH** | 22571 26.0 95% | 24447 22.0 96% | 23489 14.6 92% | 14212 7.1 79% |
| **Wellcome** | 1950 33.8 97% | 2208 34.9 96% | 1363 21.7 95% | 535 9.3 80% |

**Table 3.** Descriptive statistics for the Wikipedia URL count data. Each set contains three fields: Medicine; Biochemistry, Genetics and Molecular Biology; and Immunology and Microbiology. When the MEDI set is excluded from 2016, as needed for some formulae, the sample sizes are: World 1000; MRC 690; NIH 1000; Wellcome 470.

| Sample Mean Nonzero | 2013 | 2014 | 2015 | 2016 |
|---|---|---|---|---|
| **World** | 1500 0.025 2.0% | 1500 0.242 2.1% | 1500 0.009 0.6% | 1500 0.003 0.4% |
| **MRC** | 1150 0.036 2.4% | 1253 0.033 2.4% | 1190 0.016 1.3% | 768 0.025 0.7% |
| **NIH** | 1500 0.039 2.6% | 1500 0.033 2.1% | 1500 0.028 2.0% | 1500 0.011 0.8% |
| **Wellcome** | 1211 0.057 3.5% | 1186 0.066 5.0% | 1084 0.030 2.0% | 532 0.009 0.6% |

MNLCS, MNPC and EMNPC values and confidence intervals were calculated for each field and year using the above formulae (1,5,8,9,11,14). The same calculations were repeated for the Mendeley reader counts, as extracted by Webometric Analyst from the free Mendeley API, and for the Wikipedia citation counts, as extracted by Webometric Analyst via Bing API searches. Here, a Mendeley reader is a user of the social reference sharing site Mendeley (Gunn, 2013) that has registered the article within their library. Such users typically have already read, or intend to read, these articles (Mohammadi, Thelwall, & Kousha, 2016). Mendeley reader counts tend to reflect academic impact (Li, Thelwall, & Giustini, 2012;

Mohammadi, Thelwall, Haustein, & Larivière, 2015) and appear earlier than citation counts (Thelwall & Sud, in press). The Wikipedia citation count is the number of pages in Wikipedia that cite a given paper and reflects to some degree that the paper is transmitting knowledge to a wider public via the encyclopaedia (Kousha & Thelwall, in press-a; Rainie & Tancer, 2007). It is included as an example of a relevant webometric indicator. The Wikipedia citation count for each article was obtained by submitting a standardised Bing Wikipedia site-specific query for the article by name, journal name, publication year and author name, as in the following example.

```
Mendoza Villanueva Vargas "Vitamin D deficiency among medical
residents" "Endocrine Practice" 2013 site:wikipedia.org/wiki/
```

Since large scale automated Bing API queries are not free, a random sample of size 500 was taken for each group and world set in order to limit the data cost. This number was selected heuristically to be large enough to give a reasonable chance of detecting differences between groups. To assess the impact of this restricted sample size, a second data set for Wikipedia was constructed by expanding the world reference sets to 5000 each. The world reference sets have the largest influence on the MNLCS confidence limits and are therefore the logical first choice for expansion.

The data for this article was extracted approximately half way through 2016. Scopus citation counts were extracted on 29 June 2016, Mendeley reader counts were downloaded between 29 June and 1 July 2016, and the Wikipedia citation count searches were conducted on 30 June 2016 for the main set and additional data for the expanded world set was collected on 16-17 July 2016.

# 7 Results

This section describes the results from the perspective of evaluating the funders and the research questions are returned to in the discussion. Here, graphs with confidence intervals are reported rather than hypothesis tests because these reveal effect sizes (Cumming, 2012) as well as being suitable for situations were multiple comparisons are possible. For the MNLCS, the effect size is in terms of the ratio of the average logged citations per article for a group to the world average. For EMNPC and MNPC, the effect size is in terms of the ratio of the average proportion of cited articles for a group to the world average. In both cases there is not a natural choice about how large the differences should be in order to count as substantial enough to be of interest and therefore the focus here is simply on whether the differences are large enough to be unlikely to be a result of chance factors.

## 7.1 MNLCS: World field/year normalised average

Each organisation tends to fund research that is more highly cited than the world average for the field and year (Figure 1). Both Wellcome and MRC also tend to fund research that is more highly cited than that of NIH, although the difference is not statistically significant for the most recent year, 2016.

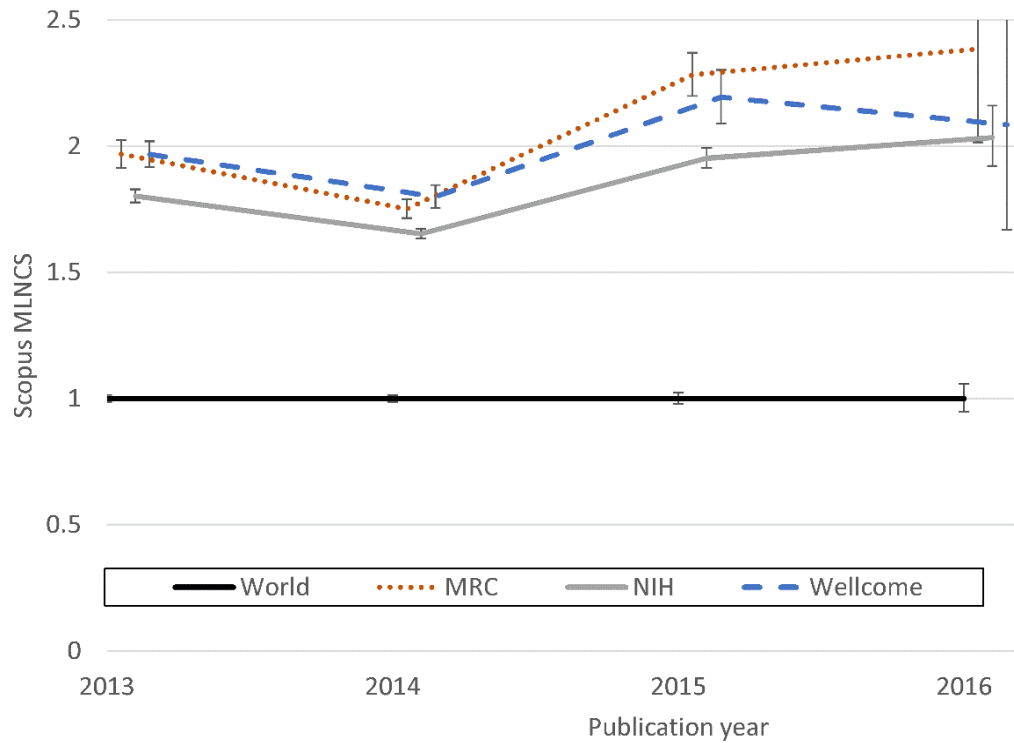

Figure 1. MNLCS for Scopus citation counts for articles funded by MRC, NIH and Wellcome in the Scopus categories BIOC, IMMU and MEDI. Funder lines are horizontally offset so that error bars do not overlap. The data is from 29 June 2016.

Each organisation tends to fund research that is more read by Mendeley users than the world average for the field or year (Figure 2). Both Wellcome and MRC also tend to fund research that is more read than that of NIH. Wellcome-funded research has significantly more readers than that of MRC in both 2014 and 2015, but they have similar numbers of readers in 2016 and the difference is at the margins of statistical significance in 2013.

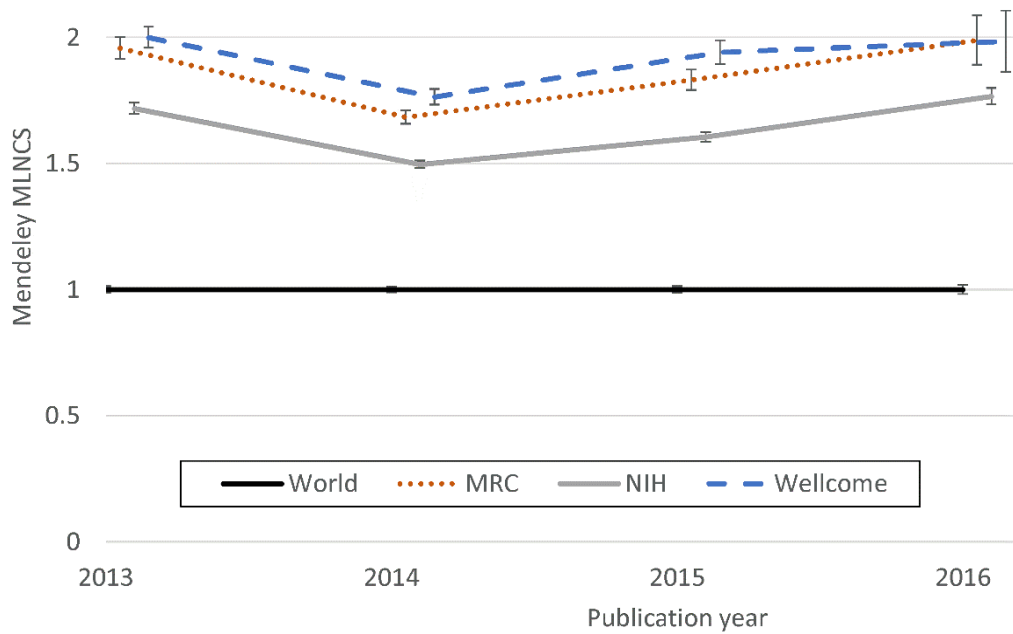

Figure 2. MNLCS for Mendeley reader counts for articles funded by MRC, NIH and Wellcome in the categories BIOC, IMMU and MEDI. Funder lines are horizontally offset so that error bars do not overlap. The data is from 29 June- 1 July 2016.

Each organisation tends to fund research that is more cited in Wikipedia than the world average in 2013 and Wellcome also in 2014 (Figure 3), suggesting that they all fund research that helps transfer knowledge to a wider public via Wikipedia. The same may be true for articles published in 2016 but the evidence is not statistically significant.

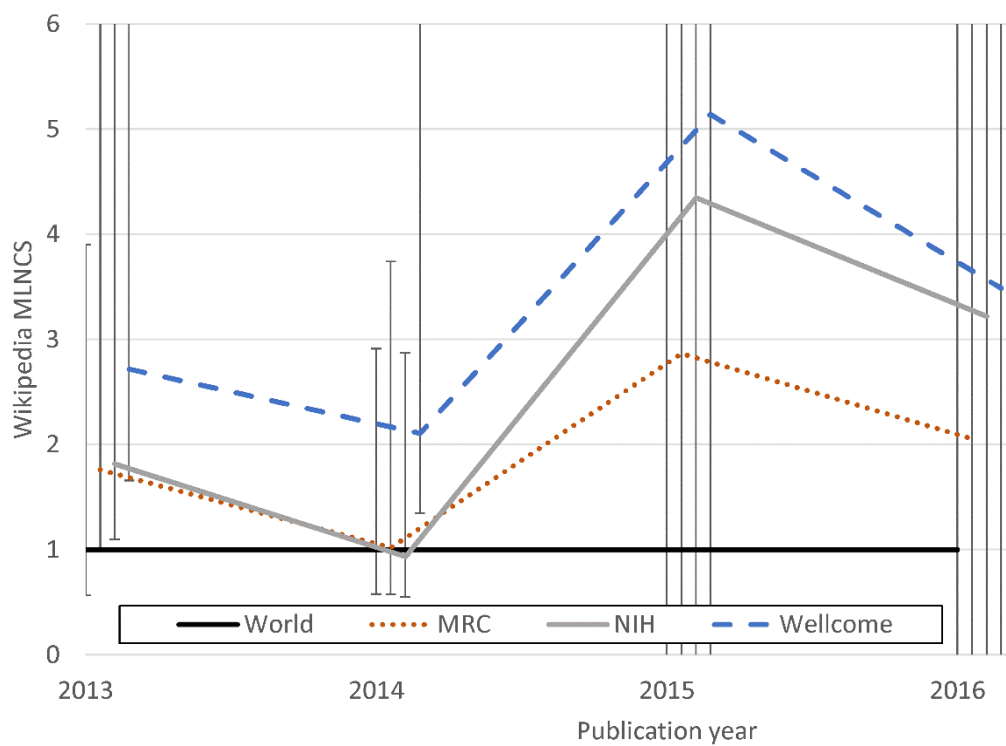

Figure 3. MNLCS for Wikipedia citation counts for articles funded by MRC, NIH and Wellcome in the categories BIOC, IMMU and MEDI. Funder lines are horizontally offset so that error bars do not overlap. A maximum of 500 articles were randomly sampled from each field/year/group combination. The 2016 MEDI category was excluded because the world set contained no matches and so field normalisation was not possible for it. Confidence limits are infinite for 2015 and 2016. The data is from 29 July 2016. The MEDI set from 2016 is excluded from the data due to a world average of 0. Confidence intervals are only approximate and are likely to be optimistic overall (see Appendix B).

The influence of the relatively small world sample can be seen from the much wider approximate confidence intervals in Figure 3 compared to Figure 4, which uses the same Wellcome, MRC and NIH data, but uses an expanded random sample of 5000 world articles for each set. The rank order between the three funders is different in 2016 because the 2016 MEDI category was excluded from the 500 article set due to the world set in this category for Figure 3 containing no matches.

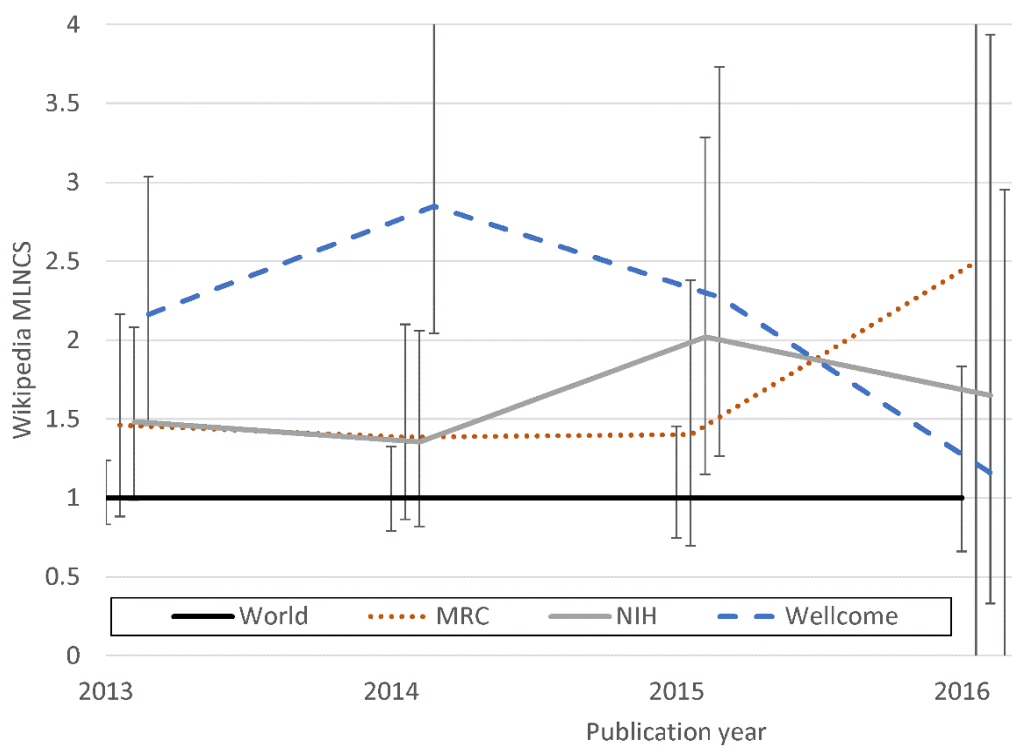

Figure 4. MNLCS for Wikipedia citation counts for articles funded by MRC, NIH and Wellcome in the categories BIOC, IMMU and MEDI. Funder lines are horizontally offset so that error bars do not overlap. A maximum of 500 articles were randomly sampled from each field/year/group combination, except for the world sets, for which a maximum of 5000 articles were randomly sampled. The group sets use the same data as for the previous figure. The group data is from 29 July 2016 and the world data is from 16-17 July 2016. Confidence intervals are only approximate and are likely to be optimistic overall (see Appendix B).

The confidence intervals for MNLCS values for each funder, using (1) and (5) on the combination of all three fields and four years, do not overlap when the data is Scopus citations or Mendeley readers (Table 4). When all the group/field combinations are restricted to 500 articles then the Wikipedia confidence intervals cannot be calculated because the variability of the world average is too large for some article sets (in 2015 and 2016). For the larger world sample size of 5000 articles per set (but a maximum of 500 articles for the group), the approximate confidence intervals can be calculated but are relatively wide (and optimistic: see Appendix B). Nevertheless, the intervals are narrow enough to exclude 1 for NIH and Wellcome, suggesting that both attract more Wikipedia citations overall than the world average. This cannot be confirmed because of the problems discussed in Appendix B.

The MNLCS confidence intervals for the world set are provided to show the stability of this figure (Table 4), even though it is, by definition always exactly 1. A wide confidence interval for a world set is likely to lead to wide confidence intervals for all group MNLCS, because world set instability translates to denominator instability in the MNLCS calculations for all groups.

**Table 4**. MNLCS for 2013-2016 and BIOC, IMMU and MEDI combined, together with sample 95% confidence intervals. The MEDI set from 2016 is excluded from the data due to a world average of 0. The MEDI set from 2016 is excluded from the Wiki 500 data due to a world average of 0. Wikipedia confidence intervals are only approximate and are likely to be optimistic overall (see Appendix B).

| Group | N | Scopus MNLCS | Mendeley MNLCS | Wiki 500 MNLCS | Wiki 500/5k MNLCS |
|---|---|---|---|---|---|
| World | 119693 | 1.000 (0.985, 1.016) | 1.000 (0.993, 1.007) | 1.000 (-,-) | 1.000 (0.871, 1.230) |
| MRC | 8107 | 2.004 (1.954, 2.054) | 1.811 (1.790, 1.833) | 1.898 (-,-) | 1.606 (0.954, 2.489) |
| NIH | 84707 | 1.840 (1.813, 1.867) | 1.631 (1.620, 1.643) | 2.521 (-,-) | 1.627 (1.190, 2.249) |
| Wellcome | 6044 | 1.968 (1.916, 2.021) | 1.900 (1.876, 1.924) | 3.290 (-,-) | 2.272 (1.817, 2.907) |

## 7.2 *EMNPC: World field/year normalised proportion of cited articles*

For Scopus citations, the three funders all have proportions of cited articles that are substantially and statistically significantly above the world average for all four years (Figure 5). Although there are differences between them in EMNPC values, these are not large except for 2015 and 2016. This is because, for all three funders, a similarly high proportion of articles had attracted citations after two years. Hence EMNPC is less discriminatory than MNLCS for older articles.

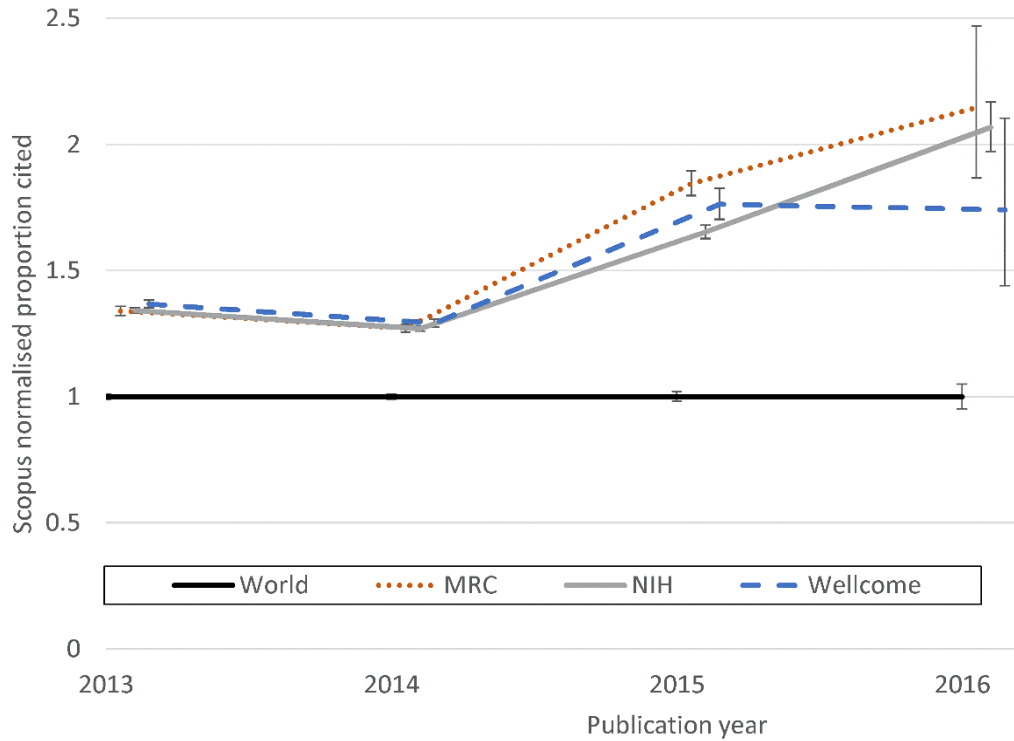

Figure 5. EMNPC for Scopus citation for articles funded by MRC, NIH and Wellcome in the Scopus categories BIOC, IMMU and MEDI. Funder lines are horizontally offset so that error bars do not overlap. The data is from 29 July 2016. Confidence intervals are approximate and may be slightly too wide (see Appendix C).

For Mendeley readers (Figure 6), the situation is very similar to that for Scopus citations. The main difference is that the groups are virtually indistinguishable in 2015 and there are only small differences between them in 2016. This is consistent with Mendeley readers appearing earlier than Scopus citations and so a shorter time is needed for a high proportion of each group's papers to have attracted at least one reader.

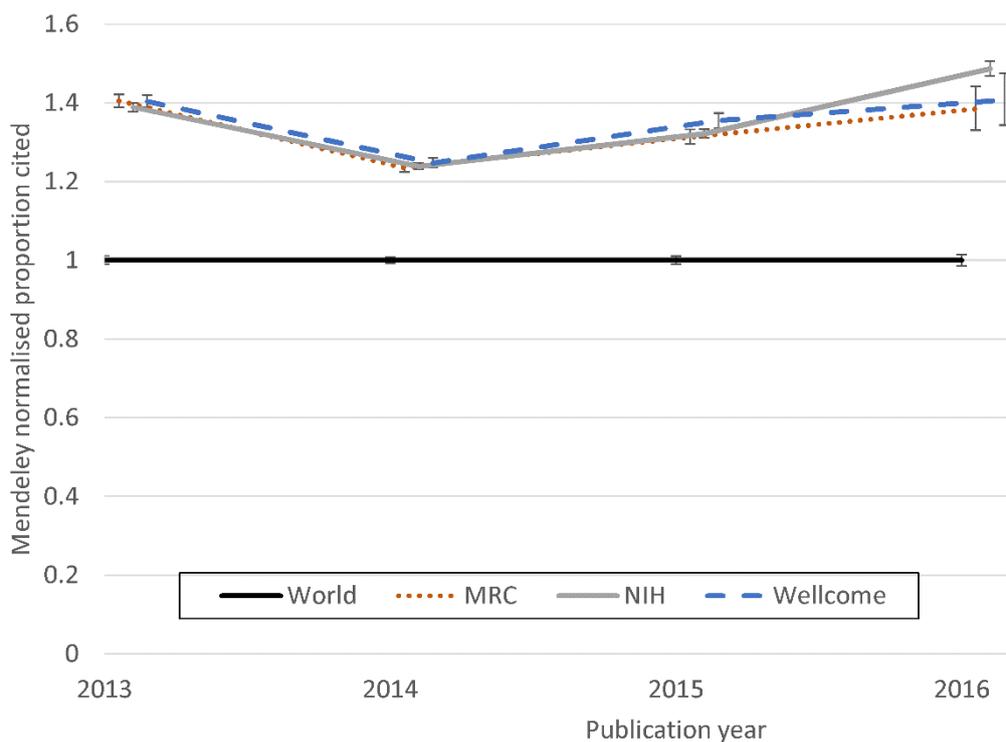

Figure 6. EMNPC for Mendeley readers for articles funded by MRC, NIH and Wellcome in the Scopus categories BIOC, IMMU and MEDI. Funder lines are horizontally offset so that error bars do not overlap. The data is from 29 June- 1 July 2016. Confidence intervals are approximate and may be slightly too wide (see Appendix C).

EMNPC values are unable to distinguish between the groups and the world average for a small majority of individual years on the Wikipedia citations 500/500 data set (Figure 7), although there are five exceptions out of 12 (Wellcome 2013-2015; NIH 2015; MRC 2016). The confidence limits for Wellcome in 2014 are also narrow enough to distinguish it from both NIH and MRC.

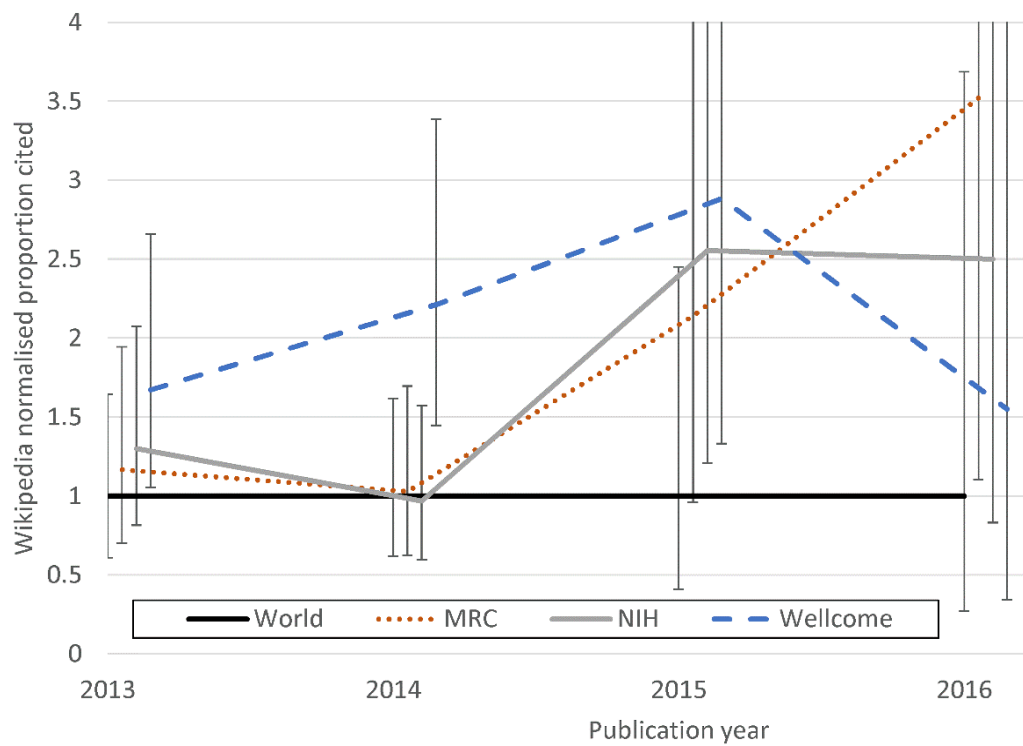

Figure 7. EMNPC for Wikipedia citations for articles funded by MRC, NIH and Wellcome in the Scopus categories BIOC, IMMU and MEDI. Funder lines are horizontally offset so that error bars do not overlap. A maximum of 500 articles were randomly sampled from each field/year/group combination. The data is from 30 June 2016. The MEDI set from 2016 is excluded due to a world average of 0. Confidence intervals are approximate and may tend to be too wide (see Appendix C).

EMNPC values are mostly able to distinguish between the groups and the world average for individual years on the Wikipedia citations 500/5000 data set, although there are two exceptions out of 12 (MRC 2013; Wellcome 2016). The confidence limits for Wellcome are also narrow enough to distinguish it from MRC in 2013, and from both NIH and MRC in 2014. Unsurprisingly, therefore, the expansion of the world data set from 500 to 5000 increased the EMNPC stability.

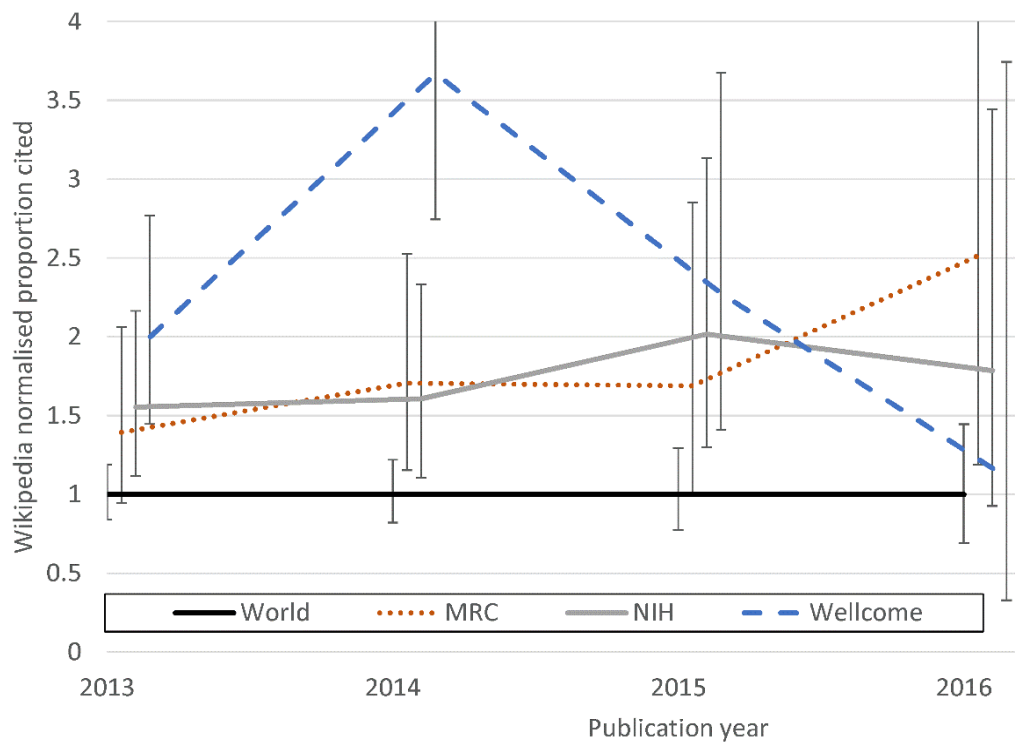

Figure 8. EMNPC for Wikipedia citations for articles funded by MRC, NIH and Wellcome in the Scopus categories BIOC, IMMU and MEDI. Funder lines are horizontally offset so that error bars do not overlap. A maximum of 500 articles were randomly sampled from each field/year/group combination, except for the world sets, for which a maximum of 5000 articles were randomly sampled. The group data is from 30 June 2016 and the world data is from 16-17 July 2016. Confidence intervals are approximate and may tend to be too wide (see Appendix C).

Combining all fields and years into a single EMNPC calculation (8 and 9), the ratio confidence intervals for each group (Table 5) do not overlap with those of the world for Scopus citations and for Mendeley readers. The confidence intervals do not contain 1 for two of the three groups for Wikipedia citations, even when there is a maximum of 500 articles per set, and the confidence intervals do not overlap when the world set is expanded to 5000 articles. Thus, overall, EMNPC seems to be more powerful than MNLCS for smaller samples. For reference, Table 6 shows the original proportions before world normalisation. Statistically significant differences are also evident here for Wikipedia citations even though only 1% of the world's articles had received any.

**Table 5**. EMNPC (sample equalised world normalised proportion of cited articles) for 2013-2016 and BIOC, IMMU and MEDI combined, together with 95% confidence intervals.

| All | N | Scopus EMNPC | Mendeley EMNPC | Wiki 500 EMNPC | Wiki 500/5k EMNPC |
|---|---|---|---|---|---|
| World | 119693 | 1.000 (0.992, 1.008) | 1.000 (0.995, 1.006) | 1.000 (0.728, 1.373) | 1.000 (0.895, 1.118) |
| MRC | 8107 | 1.463 (1.441,1.485) | 1.329 (1.318, 1.340) | 1.350 (0.982, 1.856) | 1.649 (1.299, 2.095) |
| NIH | 84707 | 1.419 (1.408, 1.429) | 1.348 (1.343, 1.354) | 1.373 (1.023, 1.844) | 1.678 (1.364, 2.062) |
| Wellcome | 6044 | 1.443 (1.418,1.468) | 1.347 (1.336, 1.359) | 2.043 (1.522, 2.741) | 2.495 (2.030, 3.066) |

**Table 6**. Sample equalised proportion of cited articles for 2013-2016 and BIOC, IMMU and MEDI combined, together with Wilson's score interval 95% confidence intervals.

| All | N | Scopus > 0 | Mendeley > 0 | Wiki 500 > 0 | Wiki 500/5k > 0 |
|---|---|---|---|---|---|
| World | 119693 | 0.487 (0.484, 0.490) | 0.679 (0.677, 0.682) | 0.013 (0.010, 0.016) | 0.010 (0.009, 0.011) |
| MRC | 8107 | 0.713 (0.703, 0.722) | 0.903 (0.896,0.909) | 0.017 (0.013, 0.021) | 0.017 (0.013, 0.021) |
| NIH | 84707 | 0.691 (0.688, 0.694) | 0.916 (0.914, 0.918) | 0.017 (0.014, 0.021) | 0.017 (0.014, 0.021) |
| Wellcome | 6044 | 0.703 (0.691,0.714) | 0.915 (0.908, 0.922) | 0.026 (0.021, 0.031) | 0.026 (0.021, 0.031) |

## 7.3  MNPC

For Scopus, the MNPC graph (Figure 9) has a slightly different shape to the EMNPC graph (Figure 5) for the Wellcome line due to the different averaging mechanisms, but otherwise the graphs are similar. The confidence intervals tend to be wider for MNPC graphs, although the difference is not large. For example, in 2015, Wellcome is indistinguishable from the other two in Figure 9 but not in Figure 5.

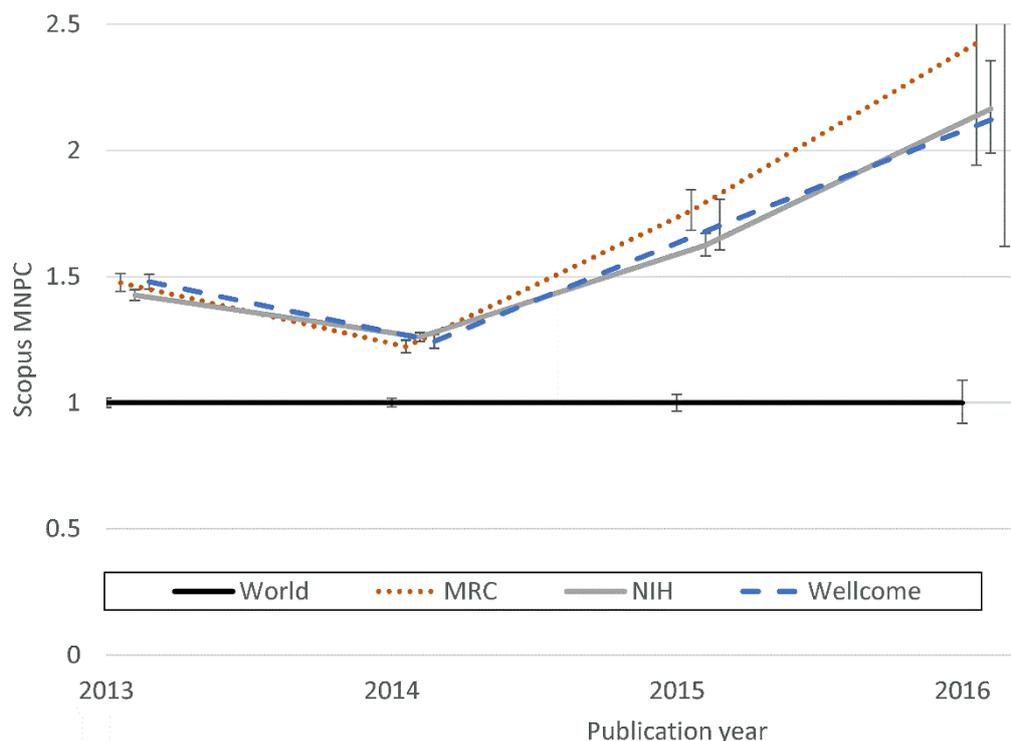

Figure 9. MNPC for Scopus citation for articles funded by MRC, NIH and Wellcome in the Scopus categories BIOC, IMMU and MEDI. Funder lines are horizontally offset so that error bars do not overlap. The data is from 29 July 2016. Confidence intervals are approximate and may tend to be too wide (see Appendix C).

For Mendeley (Figure 10), the overall shape is similar but not identical to that of the corresponding EMNPC graph (Figure 6) and the MNPC confidence intervals are again wider.

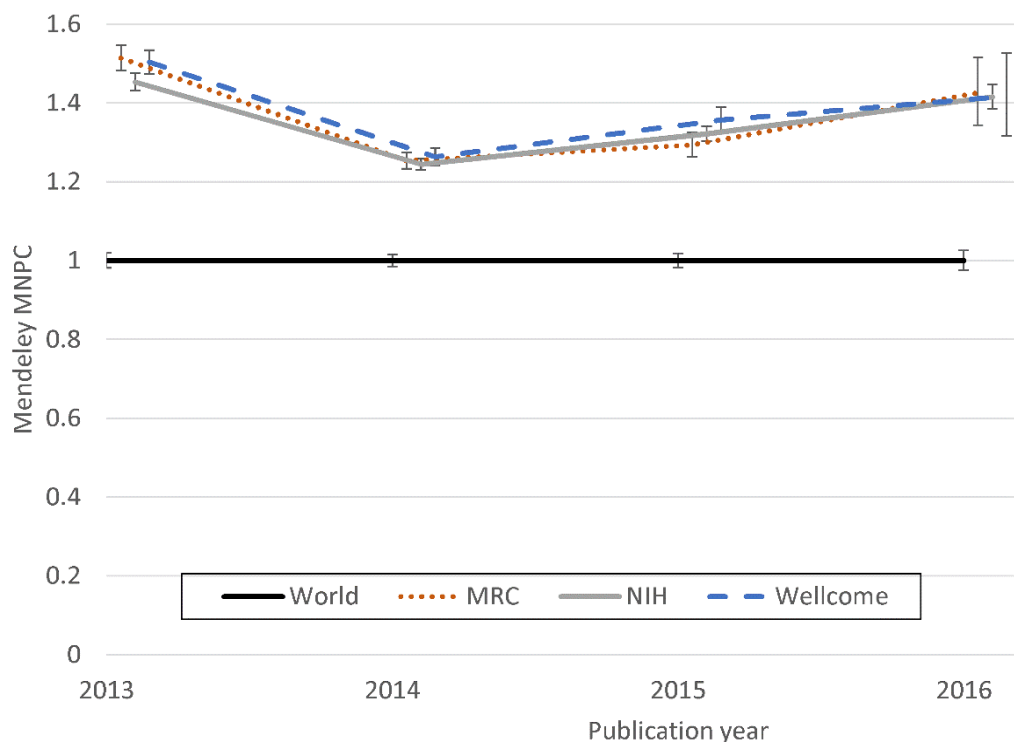

Figure 10. MNPC for Mendeley readers for articles funded by MRC, NIH and Wellcome in the Scopus categories BIOC, IMMU and MEDI. Funder lines are horizontally offset so that error

bars do not overlap. The data is from 29 June- 1 July 2016. Confidence intervals are approximate and may tend to be too narrow (see Appendix C).

For the Wikipedia citations 500/500 data set (Figure 11), the overall shape diverges from that of the corresponding EMNPC graph (Figure 7) in 2016 partly due to omitted MEDI data set for 2016 because of the divide by zero in the world set. The MNPC confidence intervals are again wider and as a result some of the differences between groups are not statistically significant in Figure 11, despite being statistically significant in Figure 7.

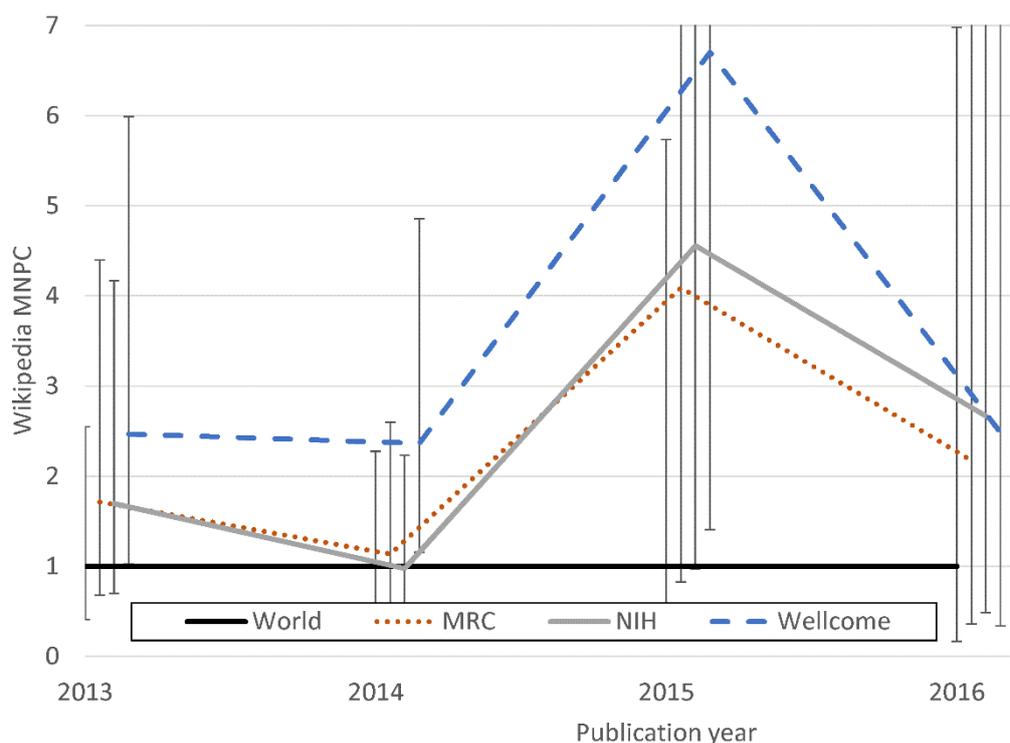

Figure 11. MNPC for Wikipedia citations for articles funded by MRC, NIH and Wellcome in the Scopus categories BIOC, IMMU and MEDI. Funder lines are horizontally offset so that error bars do not overlap. A maximum of 500 articles were randomly sampled from each field/year/group combination. The data is from 30 June 2016. The MEDI set from 2016 is excluded due to a world average of 0. Confidence intervals are approximate and may tend to be too narrow (see Appendix C).

For the Wikipedia citations 500/5000 data set (Figure 12), the Wellcome line diverges from that of the corresponding EMNPC graph (Figure 8) in 2016 due to the wider data variability. The MNPC confidence intervals are also mostly wider.

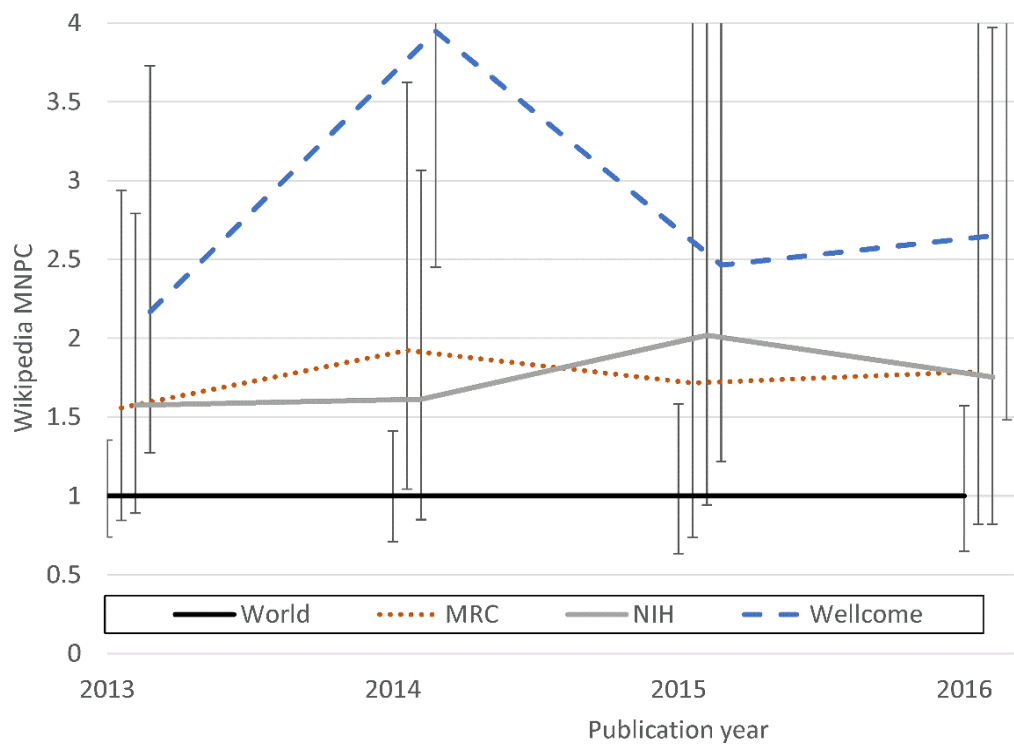

Figure 12. MNPC for Wikipedia citations for articles funded by MRC, NIH and Wellcome in the Scopus categories BIOC, IMMU and MEDI. Funder lines are horizontally offset so that error bars do not overlap. A maximum of 500 articles were randomly sampled from each field/year/group combination, except for the world sets, for which a maximum of 5000 articles were randomly sampled. The group data is from 30 June 2016 and the world data is from 16-17 July 2016. Confidence intervals are approximate and may tend to be too narrow (see Appendix C).

Combining all fields and years the confidence intervals tend to be wider for MNPC (Table 7) than those for EMNPC (Table 5). Whilst the differences are moderate for Scopus and Mendeley, they are large for the two Wikipedia data sets and result in three additional cases of group confidence intervals containing the world average (1). These three cases are NIH (both Wikipedia data sets) and MRC for the Wikipedia citations 500/5000 data set. Thus, overall, MNPC is inferior to EMNPC in terms of stability. The difference is particularly substantial for the Wikipedia data sets, presumably due to the high proportion of zeros in them. This conclusion is specific to the data sets analysed and subject to the hypothesis that the underlying advantage for each group is the same for all fields.

**Table 7**. MNPC for 2013-2016 and BIOC, IMMU and MEDI combined, together with 95% confidence intervals. A dash indicates that the calculation includes a divide by zero because the world proportion cited is zero for at least one set. The MEDI set from 2016 is excluded from the Wiki 500 data due to a world average of 0.

| All | N | Scopus MNPC | Mendeley MNPC | Wiki 500 MNPC | Wiki 500/5k MNPC |
|---|---|---|---|---|---|
| World | 119693 | 1.000 (0.963, 1.039) | 1.000 (0.981, 1.019) | 1.000 (0.320, 4.146) | 1.000 (0.651, 1.573) |
| MRC | 8107 | 1.539 (1.454, 1.639) | 1.335 (1.304, 1.368) | 2.279 (0.616, 10.076) | 1.789 (0.819, 4.360) |
| NIH | 84707 | 1.558 (1.506, 1.613) | 1.350 (1.330, 1.371) | 2.456 (0.657, 10.669) | 1.754 (0.819, 3.972) |
| Wellcome | 6044 | 1.502 (1.415, 1.609) | 1.375 (1.342, 1.410) | 3.600 (1.087, 14.634) | 2.661 (1.481, 5.176) |

## 7.4   Comparison with MNCS bootstrapped confidence intervals

For comparison with the above results, MNCS values were calculated for the same data sets (Table 8) using bootstrapping to estimate confidence intervals. Following standard MNCS practice (although not a necessary assumption) the bootstrapping assumed that the world figure of 1 is exact (i.e., bootstrapping the world normalised data rather than also bootstrapping for the world normalisation calculation). Although this extra assumption would tend to narrow the confidence intervals, they are still wider for Scopus and Mendeley than for the MNLCS values in Table 4. For example, the Wellcome Scopus MNLCS confidence interval width is 2.021-1.916=0.105 whereas the Wellcome Scopus MNCS confidence interval width is nearly five times wider 3.396-2.899=0.497. Even after compensating for the larger MNCS values, the confidence intervals are still substantially wider than for MNLCS (e.g., 0.497/3.132=1.16 rather than 0.105/1.968=0.05 for Wellcome). This confirms that the MNLCS formula can give narrower confidence intervals than MNCS bootstrapping, despite relaxing the assumption that the world average is exact. Thus, the confidence intervals in this case are both more theoretically robust and narrower. The theoretical robustness would be irrelevant for very large sample sizes, but the confidence intervals should still be narrower for the MLNCS.

For the Wikipedia 500 data sets, MNCS confidence intervals can be calculated, in contrast to the situation for MNLCS, due to the assumption that the world mean is exact for standard MNCS calculations. For the Wikipedia 500/5k data sets, the confidence intervals have similar relative widths between the MNCS and MNLCS. In both of these cases, however, the high variability of the Wikipedia citation data makes the world averages unreliable and so the MNCS confidence intervals are also unreliable. This is under the theoretical assumption that the world set is a sample of the possible articles that could have been produced under similar situations, thereby treating the population as a sample, as discussed at the end of Section 2.

**Table 8**. MNCS for 2013-2016 and BIOC, IMMU and MEDI combined, together with bootstrapped 95% confidence intervals.

| All | Scopus > 0 | Mendeley > 0 | Wiki 500 > 0 | Wiki 500/5k > 0 |
|---|---|---|---|---|
| World | 1.000 | 1.000 | 1.000 | 1.000 |
| | (0.969, 1.036) | (0.988, 1.011) | (0.659, 1.407) | (0.823, 1.198) |
| MRC | 2.867 | 2.856 | 1.703 | 1.506 |
| | (2.760, 2.979) | (2.764, 2.955) | (1.176, 2.314) | (0.624, 3.086) |
| NIH | 2.313 | 2.255 | 2.836 | 1.311 |
| | (2.278, 2.352) | (2.226, 2.286) | (1.779, 4.109) | (0.890, 1.823) |
| Wellcome | 3.132 | 3.308 | 3.231 | 1.421 |
| | (2.899, 3.396) | (3.131, 3.518) | (2.270, 4.416) | (1.115, 1.748) |

# 8   Discussion

The main limitation of this study is that only three groups were investigated (MRC, NIH, Wellcome) and that the empirical results are therefore not conclusive. They primarily serve to illustrate the new methods introduced, show that the claims are broadly credible, and demonstrate that the formulae are capable of generating useful results. A practical limitation of the indicators is that some subject categories in Scopus and WoS contain periodicals that are rarely cited (e.g., trade publications and arts magazines) and these categories therefore have unusually high numbers of uncited articles (Thelwall, 2016c). This can inflate EMNPC, MNPC and MNLCS values for sets of articles that do not include any rarely cited periodical articles by reducing the world category average. It is therefore important for world normalisation purposes to ensure that such periodicals are identified and excluded from the data.

Most of the graphs show a dip in the lines in 2014 for all three funders. Since it seems unlikely that all three funded less impactful research in the same year, the most likely cause is that the world average increased relative to them in 2014. This could be through the publication of a batch of highly cited articles from another funder but it may also be due to a technical issue, such as a change in the journal composition of one of the three Scopus categories in 2014 by removing low-cited journals that do not publish much research funded by Wellcome, NIH and MRC. For example, these might be non-English journals since international coverage of citation databases is uneven (de Moya-Anegón, Chinchilla-Rodríguez, Vargas-Quesada, et al., 2007).

As mentioned above, a limitation for the specific example here is that Scopus acknowledgements seem to index only about a third of Wellcome-funded papers and if this is due to primarily indexing acknowledgements in high impact journals then the world normalised comparisons would be unfair. An important limitation from the perspective of the validity of the results is that there are many different reasons why each organisation apparently funds articles with relatively high indicator values. These results could be due to the organisations being successful in selecting excellent research to fund, to researchers receiving funding being able to conduct better research in consequence of the funding, to the organisations funding high citation sub-specialisms, or to the funded research having a higher profile due to funder publicity. Thus, practical interpretations of the graphs and tables require detailed knowledge about the types of research that they fund and their publicity policies (e.g., open access mandates, press release strategies).

In answer to the first research question, the MNLCS was able to distinguish statistically between the different groups assessed and the world average for most individual years and overall for Scopus citations and Mendeley reader counts. It was much less discriminatory for Wikipedia citations, being unable to distinguish between groups and the world average on the 500/500 data set (using the approximate confidence intervals), except for Wellcome in 2013 and 2014. It was more discriminatory on the 500/5000 data set, but still with relatively wide confidence intervals and not for the MRC. Nevertheless, since the MNLCS confidence intervals are unreliable for the Wikipedia citations (Appendix B), these conclusions are not statistically robust. For the normalised proportion cited, the EMNPC values for the Mendeley and Scopus citations were discriminatory overall and for most individual years for Scopus citations, Mendeley readers and Wikipedia citations from the 500/5000 data set (although the results are not robust for Wikipedia). EMNPC and MNPC do not seem to be very useful to distinguish one group from another for Scopus citations or Mendeley readers, except perhaps for the most recent year, because of the high proportion of cited or read articles. This may be different for non-medical subjects with lower citation rates. For the Wikipedia citations from the 500/500 data set, EMNPC values are sufficient to distinguish all groups from the world average overall and for some individual years, showing that EMNPC is partially effective in this low data case. MNPC is not quite as stable but still represents an advantage over MNLCS. As a reminder, this conclusion depends upon the data sets analysed and the hypothesis that the underlying advantage for each group is the same for all fields.

In answer to the second research question, the data sets for which the proportion of cited articles was low were the two Wikipedia citation data sets. As the above paragraph argues, EMNPC and MNPC are preferable to MNLCS in these contexts for more discriminatory power as well as more robust confidence intervals.

The comparison between EMNPC and MNPC suggests that EMNPC is superior for stability and also has the advantage of being able to use all data, whereas the MNPC calculation (and MNCS, MNLCS) had to exclude the MEDI 2016 Wikipedia 500/500 set because none of the world's articles were cited. Hence the data *for this paper* suggests that EMNPC is preferable to MNPC in all respects. The cause of the reduced stability of MNPC is that, as a weighted sum of ratios, if any of the ratios in the weighted sum has a wide confidence interval then this has a substantial effect on the overall confidence interval. Nevertheless, MNPC has two advantages that the current data sets have not revealed: it is not restricted to sets of fields of approximately equal sizes and therefore for some data sets can have more complete coverage than EMNPC, and it is fairer because each article is weighted exactly equally, whereas the EMNPC calculation gives higher weightings to articles in smaller fields. Thus, MNPC is recommended for cases where field sizes are unequal and it is important not to exclude any, especially when none of the world groups has few uncited articles so that the confidence intervals are not large.

A limitation of the method is that the heuristic confidence intervals introduced for both MNPC and EMNPC are estimates and the MNPC formula estimates differ substantially from MNPC estimates from bootstrapping (Appendix C). A simulation approach is needed to assess which of the two is most accurate (simulating datasets with a range of different parameters and comparing bootstrapping confidence intervals, confidence intervals from runs of the model, and confidence intervals from the formula).

An important omission in the current work is for guidelines to select when field/year combinations contain too few articles relative to other field/year combinations, so that they should be excluded from EMNPC calculations. Future work is needed to investigate this issue.

# 9 Conclusions

This article introduced new field (and year) normalised formulae for indicators, MNLCS, MNPC and EMNPC, all of which can be used to generate estimates of how far above or below the world average a set of articles is. It is possible to calculate approximate confidence intervals for them without bootstrapping as long as the raw data approximately follows the discretised lognormal distribution (MNLCS) and does not have too many zeros (MNPC), and so they seem to be superior to many previous indicators for citations in this regard (although not percentile indicators). Nevertheless, for the MNLCS and MNPC the confidence intervals are estimates and for EMNPC the calculation relies upon the assumption that the group advantage or disadvantage of the set of articles assessed is the same in all fields and years. The MNLCS indicator is the preferred option, when it is applicable, because it exploits more information – the exact citation count for each article – and therefore has the potential to be more stable. The EMNPC is recommended for cases where few articles are cited, as occurs for many alternative indicators. The EMNPC is preferable in these cases because it impossible to construct narrow or robust confidence intervals for the MNLCS when most of the data consists of zeros. The MNPC formula is more suitable than EMNPC when there are small field/year groups and there are no wide confidence intervals in the MNPC calculation. Note that these conclusions are consistent with the analytical arguments presented in this paper and the medical funders example provided but still need to be evaluated with a range of different examples to verify them.

Based on the arguments above and consistent with the example assessed in this article, a practical overall strategy for calculating a set of field normalised alternative indicators for a group of publications is now possible for the first time and is supported by functions added to the free software Webometric Analyst (http://lexiurl.wlv.ac.uk/) for this article (see also the guidelines here: Thelwall, 2017).

1. Identify the group of publications to be assessed and categorise them by field (e.g., using Scopus or WoS subject categories).
2. Save the article information (authors, title, journal, publication year) in a standard tab-delimited format in a separate file for each subject category/year combination. Discard publications that are in small subject/year combinations (e.g., <100 publications).
3. For each retained subject/year combination, download all articles from Scopus/WoS (if possible) or a large balanced sample (e.g., the first and last 5000 articles published in the category) for the world reference set. Filter out any large trade or art journals with a high proportion of uncited articles. Name the files using the standard Webometric Analyst naming convention (see http://lexiurl.wlv.ac.uk/).
4. Decide which alternative indicators are to be used for the data.
5. For free alternative indicators (e.g., Mendeley readers). Use Webometric Analyst to download all indicator values, or use another altmetric data source if available.
6. For paid alternative indicators, use Webometric Analyst to generate a random sample of articles from the world and group sets (e.g., 500 per set) and use these samples instead of the full set.
7. Use Webometric Analyst to calculate MNLCS, MNPC and EMNPC values and confidence limits for both. These values can be calculated separately for each year, combined across all years, or both. MNLCS is recommended as the main indictor unless the proportions cited are low, in which case EMNPC or MNPC are preferred. MNPC is preferable to EMNPC if it is stable enough. This is likely to occur for very large sample sizes or proportions cited that are not too low. If MNPC is used, then the its confidence intervals should be treated with caution.

Finally, the results in this article also give new evidence that Mendeley reader counts give earlier evidence of impact than do Scopus citations (e.g., see: Fairclough & Thelwall, 2015ab), and also show, for the first time, that web-based indicators can be used to assess whether a group of articles has had a type of impact that is significantly different from the world average.

# Appendix A: Indicator calculation examples

This section gives a tiny example of the MNLCS calculations. Here a group publishes 5 articles in field A and 5 in field B (Table A1) and the rest of the world publishes an additional 5 articles in field A and 5 in field B (Table A2). All calculations are performed to full calculator accuracy but reported to two decimal places. The first and fourth columns in each table give the raw citation counts and the second and fifth columns give the log transformed citation counts. The following paragraph explains the third and sixth columns.

For normalisation, world average log citation counts are needed. The world average log citation count for field A is the average number of log citations to all articles in field A, which are the 5 articles from the group (Table A1 left hand side) and the 5 from the rest of the world (Table A2 left hand side). The world average log citation count for field A is therefore (4.19+2.20)/10=0.64 (2 DP). The normalised citation count for all articles in field A is therefore Ln(1+ citations)/0.64. Similarly, the world average log citation count for field B is (3.58+4.68)/10=0.83. The normalised citation count for all articles in field B is therefore Ln(1+ citations)/0.83.

Ignoring Field B, the MNLCS value for the group within field A alone is the arithmetic mean of the normalised log citations (Ln(1+ citations)/0.64), which is 6.56/5=1.31. Similarly, the MNLCS value for the group within field B alone is 4.34/5=0.87. For field A, the world MNLCS is the average of the normalised log citations, or (6.56+3.44)/10=1, as expected. Similarly, for field B world MNLCS is (4.34+5.66)/10=1, as expected.

For the complete set of publications, the group MNLCS is the average of all normalised log citations, (6.56+4.34)/10=1.09. The world MNLCS is (6.56+3.44+4.34+5.66)/20=1, as again expected.

**Table A1**. Artificial sample of 5 articles in field A and 5 articles in field B published by a research group.

| Field A | | | Field B | | |
|---|---|---|---|---|---|
| Citations | Ln(1+ citations) | Ln(1+ citations)/ 0.64 | Citations | Ln(1+ citations) | Ln(1+ citations)/ 0.83 |
| 0 | 0 | 0 | 0 | 0 | 0 |
| 0 | 0 | 0 | 1 | 0.69 | 0.84 |
| 1 | 0.69 | 1.09 | 1 | 0.69 | 0.84 |
| 2 | 1.10 | 1.72 | 2 | 1.10 | 1.33 |
| 10 | 2.40 | 3.75 | 2 | 1.10 | 1.33 |
| **Sum** | **4.19** | **6.56** | **Sum** | **3.58** | **4.34** |

**Table A2**. Artificial sample of 5 articles in field A and 5 articles in field B published by the rest of the world.

| Field A | | | | Field B | | |
|---|---|---|---|---|---|---|
| Citations | Ln(1+ citations) | Ln(1+ citations)/ 0.64 | | Citations | Ln(1+ citations) | Ln(1+ citations)/ 0.83 |
| 0 | 0 | 0 | | 0 | 0 | 0 |
| 0 | 0 | 0 | | 1 | 0.69 | 0.84 |
| 0 | 0 | 0 | | 2 | 1.10 | 1.33 |
| 2 | 1.10 | 1.72 | | 2 | 1.10 | 1.33 |
| 2 | 1.10 | 1.72 | | 5 | 1.79 | 2.17 |
| **Sum** | **2.20** | **3.44** | | **Sum** | **4.68** | **5.66** |

The EMNPC calculations for the same example are given in Table A3. Here the sample sizes are already equal so no calculations are needed for the equalisation of sample sizes.

**Table A3**. EMNPC calculations for field A, field B and overall (combining fields A and B) for the citation counts in Table A1 and A2.

| | Number cited | | | | Proportion cited | | | EMNPC | | |
|---|---|---|---|---|---|---|---|---|---|---|
| | Field A | Field B | All | | Field A | Field B | All | Field A | Field B | All |
| Group | 3 | 4 | 7 | | 0.60 =3/5 | 0.80 =4/5 | 0.70 =7/10 | 1.20 =0.60/0.50 | 1.00 =0.80/0.80 | 1.08 =0.70/0.65 |
| World | 5 | 8 | 13 | | 0.50 =5/10 | 0.80 =8/10 | 0.65 =13/20 | 1.00 =0.50/0.50 | 1.00 =0.80/0.80 | 1.00 =0.65/0.65 |

Using the figures in Table A3 and the formula above (12), the MNPC calculations can be expressed as a weighted sum of the ratios of the group proportion cited to the world proportion cited for each field.

$$\text{MNPC} = \sum_{f \in \{A,B\}} \frac{n_{gf}}{n_g} \times \frac{p_{gf}}{p_{wf}} = \frac{5}{10} \times \frac{3/5}{5/10} + \frac{5}{10} \times \frac{4/5}{8/10} = \frac{11}{10} = 1.1$$

As an additional example, suppose that another group publishes 100 articles in Field C, with 8% of them cited compared to a world average of 4% and the group also publishes 200 articles in field D with 5% of them cited compared to a world average of 20%. Using simplified formulae (12) and (13) EMNPC and MNPC can be calculated as follows.

$$\text{EMNPC} = \frac{\sum_{f \in \{C,D\}} p_{gf}}{\sum_{f \in \{C,D\}} p_{wf}} = \frac{0.08 + 0.05}{0.04 + 0.2} = \frac{0.13}{0.24} = 0.542$$

$$\text{MNPC} = \sum_{f \in \{C,D\}} \frac{n_{gf}}{n_g} \times \frac{p_{gf}}{p_{wf}} = \frac{100}{300} \times \frac{0.08}{0.04} + \frac{200}{300} \times \frac{0.05}{0.2} = 0.833$$

# Appendix B: Normal distribution formula bootstrapping tests

Table B1 reports comparisons of confidence intervals calculated with the normal distribution formula with confidence intervals calculated with 1000 bootstrapping iterations. For each iteration, a new sample of the same size as the original data set was created by randomly selecting data points from the original data set with replacement. The mean was then calculated of this artificial sample. After 1000 repetitions, a 95% bootstrap confidence interval was created by arranging the 1000 means in ascending order and selecting the means at the 2.5 and 97.5 percentiles.

The lower limit percentage difference was calculated by subtracting the width of the lower half of the 95% confidence interval (i.e., subtracting the lower limit from the mean) from bootstrapping from the width of the lower half of the 95% confidence interval, as calculated by the formula. The difference in widths was then divided by the width of the bootstrapping confidence interval to give a percentage difference. The purpose of this was to test whether it was reasonable to use the normal distribution confidence interval formula. The test is for the log-transformed data (type Ln(1+x) in Table B1), with the other values included in the table for comparison purposes. The tests were conducted 48 times for each data source, once for each year (a total of 4) as well as once for the world set and each group (4) and once for each field (3).

From Table B1, the assumption that the data sets for each field, group and year are approximately normal after the log transformation is supported for Scopus and Mendeley by the average absolute difference in confidence interval widths between the bootstrap methods and formula being 3%-4%, and the maximum difference for any of the 48 data sets being 14%. In conjunction with evidence that the discretised lognormal distribution fits both Scopus citation counts and Mendeley reader counts well (Fairclough & Thelwall, 2015b; Thelwall, 2016ab), this supports the use of the normal distribution formula. Data of these types cannot exactly fit the normal distribution because they are discrete, but can be thought of as following the normal distribution for the purpose of using the confidence interval formula. Without the log transformation, the confidence interval formula is only half as accurate and has a small systematic bias (type x in Table B1).

The confidence interval widths from the formula agree considerably less with the bootstrap confidence interval widths for the two Wikipedia data sets, with average absolute differences of up to 23%, a systematic bias (the formula interval tends to be too narrow on the lower half and too wide on the upper half), and differences in widths of up to 100%. Thus, the formulae are unreliable for the Wikipedia data and their results should therefore be interpreted with great caution. Possible reasons for the discrepancy are: (a) Wikipedia citations counts do not follow a discretised lognormal distribution; (b) Wikipedia citations follow a discretised lognormal distribution but (b1) the high number of zeros in the data means that the discretisation process breaks the connection with the continuous normal distribution to the extent that the formula does not work, or (b2) larger sample sizes are needed for the formula to be effective due to the high number of zeros.

**Table B1**. A comparison of confidence interval left and right hand side widths, as calculated with the normal distribution formula and bootstrapping with 1000 iterations. Positive numbers in the *average % difference* columns indicate that the formula confidence interval is narrower than the bootstrap confidence interval. All rows are calculated from 48 datasets.

| Data | Type | Lower limit average % difference | Upper limit average % difference | Lower limit average % absolute difference | Upper limit average % absolute difference | Lower limit max. % difference | Upper limit max. % difference |
|------|------|------|------|------|------|------|------|
| Wiki 500 500 | Ln(1+x) | 23% | -6% | 23% | 7% | 100% | 15% |
| Wiki 500 5k | Ln(1+x) | 19% | -7% | 19% | 8% | 100% | 34% |
| Scopus | Ln(1+x) | 1% | 0% | 4% | 3% | 12% | 9% |
| Mendeley | Ln(1+x) | 0% | 1% | 4% | 3% | 10% | 14% |
| Wiki 500 500 | x | 28% | -6% | 28% | 7% | 100% | 22% |
| Wiki 500 5k | x | 24% | -7% | 24% | 7% | 100% | 22% |
| Scopus | x | 7% | -4% | 9% | 5% | 39% | 20% |
| Mendeley | x | 7% | -4% | 8% | 5% | 35% | 17% |

# Appendix C: EMNPC and MNPC confidence interval formula bootstrapping tests

The confidence interval widths predicted by the EMNPC formula agree very approximately with confidence intervals calculated by bootstrapping (Table C1). For the complete set, the confidence interval for the formula is relatively optimistic in the sense of being narrower than the bootstrap confidence intervals. It is not clear which confidence interval is the most accurate.

**Table C1**. A comparison of confidence interval left and right hand side widths, as calculated with the EMNPC formula and bootstrapping with 10000 iterations. Positive numbers indicate that the formula confidence interval is narrower than the bootstrap confidence interval.

| Av. % difference | | Wiki 500 500 Low. 95 | Wiki 500 500 Upp. 95 | Wiki 500 5k Low. 95 | Wiki 500 5k Upp. 95 | Scopus Low. 95 | Scopus Upp. 95 | Mendeley Low. 95 | Mendeley Upp. 95 |
|------|------|------|------|------|------|------|------|------|------|
| Year(s) | Group | Low. 95 | Upp. 95 | Low. 95 | Upp. 95 | Low. 95 | Upp. 95 | Low. 95 | Upp. 95 |
| 2013 | MRC | -10% | -7% | -19% | -4% | -17% | -7% | -12% | -4% |
| 2013 | NIH | 1% | -5% | -8% | 10% | 2% | -2% | 0% | -1% |
| 2013 | Wellcome | -4% | -8% | -10% | 5% | 0% | 5% | -12% | -10% |
| 2014 | MRC | -5% | -4% | -11% | 12% | -18% | -13% | -20% | -12% |
| 2014 | NIH | -3% | -6% | -9% | 8% | -2% | -2% | 2% | -1% |
| 2014 | Wellcome | -2% | -9% | -12% | 1% | -10% | -7% | -15% | -10% |
| 2015 | MRC | -4% | -28% | -17% | 8% | -20% | -17% | -16% | -11% |
| 2015 | NIH | 3% | -32% | -11% | 11% | -4% | -2% | 0% | -2% |
| 2015 | Wellcome | 0% | -30% | -13% | 11% | -19% | -16% | -20% | -11% |
| 2016 | MRC | -23% | -24% | -34% | -13% | -24% | -17% | -26% | -22% |
| 2016 | NIH | -12% | -52% | -18% | 17% | -15% | -18% | -4% | -3% |
| 2016 | Wellcome | -22% | -52% | -30% | 69% | -17% | -6% | -23% | -15% |
| All | MRC | -17% | 23% | -16% | -1% | -11% | -12% | -37% | -35% |
| All | NIH | -3% | 3% | -7% | 6% | 12% | 14% | -6% | -1% |
| All | Wellcome | 1% | -4% | -4% | 7% | -3% | -3% | -38% | -37% |

The confidence interval widths predicted by the MNPC formula are substantially optimistic compared to the formulae calculated by bootstrapping (Table C2). The MNPC calculation is most relevant for the Wiki data sets because of their low proportions cited (for which the MNPC/EMNPC was designed). For one of these, Wiki 500 500, bootstrapping confidence intervals could not be calculated for 2015 onwards for at least 2.5% of the bootstrapping samples due to a divide by zero (no uncited articles in the bootstrapped world set).

**Table C2**. A comparison of confidence interval left and right hand side widths, as calculated with the MNPC formula and bootstrapping with 10000 iterations. Positive numbers indicate that the formula confidence interval is narrower than the bootstrap confidence interval.

| Av. % diff. | | Wiki 500 500 | Wiki 500 500 | Wiki 500 5k | Wiki 500 5k | Scopus | Scopus | Mendeley | Mendeley |
|---|---|---|---|---|---|---|---|---|---|
| Year(s) | Group | Low. 95 | Upp. 95 | Low. 95 | Upp. 95 | Low. 95 | Upp. 95 | Low. 95 | Upp. 95 |
| 2013 | MRC | 33% | -28% | 24% | 92% | 39% | 42% | 34% | 43% |
| 2013 | NIH | 46% | -36% | 42% | 105% | 43% | 41% | 42% | 46% |
| 2013 | Wellcome | 43% | -41% | 39% | 90% | 37% | 36% | 48% | 47% |
| 2014 | MRC | 38% | 29% | 28% | 88% | 53% | 56% | 43% | 50% |
| 2014 | NIH | 49% | 37% | 38% | 105% | 64% | 63% | 63% | 60% |
| 2014 | Wellcome | 51% | 15% | 39% | 79% | 52% | 51% | 46% | 50% |
| 2015 | MRC | 27% | - | 21% | 114% | 58% | 64% | 56% | 58% |
| 2015 | NIH | 32% | - | 33% | 103% | 64% | 64% | 61% | 57% |
| 2015 | Wellcome | 27% | - | 21% | 100% | 59% | 69% | 58% | 66% |
| 2016 | MRC | 11% | - | 0% | 134% | 44% | 69% | 59% | 66% |
| 2016 | NIH | 14% | - | 10% | 103% | 52% | 53% | 62% | 54% |
| 2016 | Wellcome | -14% | - | -25% | 62% | 37% | 76% | 56% | 75% |
| All | MRC | 106% | - | 126% | 321% | 139% | 158% | 195% | 205% |
| All | NIH | 127% | - | 146% | 273% | 152% | 143% | 199% | 211% |
| All | Wellcome | 30% | - | 153% | 246% | 131% | 169% | 204% | 206% |

# 11 Appendix D: MNPC individual field formulae

These are the formulae for lower and upper limits for MNPC for the single field $f$ (Bailey, 1987). A continuity correction of 0.5 may be added to all $pn$ terms.

$$\mathrm{MNPC}_{fL} = \exp\left( \ln\left(\frac{p_{gf}}{p_{wf}}\right) \right.$$
$$\left. - 1.96 \sqrt{\frac{(n_{gf} - p_{gf}n_{gf})/(p_{gf}n_{gf}) + (n_{wf} - p_{wf}n_{wf})/(p_{wf}n_{wf})}{n_{gf}} + \frac{(n_{wf} - p_{wf}n_{wf})/(p_{wf}n_{wf})}{n_{wf}}} \right)$$

$$\mathrm{MNPC}_{fU} = \exp\left( \ln\left(\frac{p_{gf}}{p_{wf}}\right) \right.$$
$$\left. + 1.96 \sqrt{\frac{(n_{gf} - p_{gf}n_{gf})/(p_{gf}n_{gf}) + (n_{wf} - p_{wf}n_{wf})/(p_{wf}n_{wf})}{n_{gf}} + \frac{(n_{wf} - p_{wf}n_{wf})/(p_{wf}n_{wf})}{n_{wf}}} \right)$$